\documentclass[12pt]{article}
\usepackage[paper=a4paper,margin=1in]{geometry}
\usepackage{t1enc}      

\usepackage{amsmath}
\usepackage{amsfonts}
\usepackage{amssymb}
\usepackage{amsthm}
\theoremstyle{plain}
\usepackage{graphicx}
\usepackage{mathrsfs}  

\newcommand{\ua}{\underline a \,}
\newcommand{\ub}{\underline b \,}
\newcommand{\uc}{\underline c \,}
\newcommand{\ud}{\underline d \,}

\newcommand{\bi}{\bf i}
\newcommand{\bj}{\bf j}

\newcommand{\Clim}{{\rm Clim\,}}

\newtheorem{theorem}{Theorem}[subsection]
\newtheorem{proposition}{Proposition}[subsection]
\newtheorem{lemma}{Lemma}[subsection]
\newtheorem{corollary}[theorem]{Corollary}

\newtheorem*{theorem*}{Theorem}
\newtheorem*{lemma*}{Lemma}
\newtheorem*{proposition*}{Proposition}

\providecommand{\keywords}[1]
{\small	\textbf{\textit{Keywords:}} #1 }

\numberwithin{equation}{section}

\begin{document}
\bibliographystyle{unsrt}

\title{Curved spacetimes from quantum mechanics}

\author{L\'aszl\'o B. Szabados\footnote{Budapest, European Union; e-mail:
lsodabazs(at)gmail.com}}

\maketitle

\begin{abstract}
The ultimate extension of Penrose's Spin Geometry Theorem is given. It is 
shown how the \emph{local} geometry of any \emph{curved} Lorentzian 
4-manifold (with $C^2$ metric) can be derived in the classical limit using 
only the observables in the algebraic formulation of abstract 
Poincar\'e-invariant elementary quantum mechanical systems. In particular, 
for any point $q$ of the classical spacetime manifold and curvature tensor 
there, there exists a composite system built from finitely many 
Poincar\'e-invariant elementary quantum mechanical systems and a sequence 
of its states, defining the classical limit, such that, in this limit, the 
value of the distance observables in these states tends with asymptotically 
vanishing uncertainty to lengths of spacelike geodesic segments in a convex 
normal neighbourhood $U$ of $q$ that determine the components of the 
curvature tensor at $q$. Since the curvature at $q$ determines the metric 
on $U$ up to third order corrections, the metric structure of curved $C^2$ 
Lorentzian 4-manifolds is recovered from (or, alternatively, can be 
\emph{defined} by the observables of) abstract Poincar\'e-invariant 
quantum mechanical systems. 
\end{abstract}

\keywords{curved Lorentzian geometries, sectional curvature, empirical
distance, classical limit}


\section{Introduction}
\label{sec-1}

As is well known, the notion of spacetime of special and general relativity, 
as introduced \emph{operationally} in the framework of \emph{classical 
  physics} e.g. in \cite{EPS}, turned out to be surprisingly useful even in 
\emph{quantum physics}. But Marzke and Wheeler \cite{MW}, quoting Bohr and 
Rosenfeld, stress that \emph{`... every proper theory should provide in and 
by itself its own means for defining the quantities with which it deals'}. 
Hence, in particular, the basic notions in quantum physics should not be 
based on classical physical concepts, and the laws of classical physics 
should be derivable from the notions and laws of micro-physics. The latter 
may not be based on concepts of classical physics. Extending this principle 
to the spacetime of classical general relativity as a background for the 
quantum physical calculations, it should also be re-defined by purely quantum 
concepts. In the contribution \cite{MW}, Marzke and Wheeler raise the idea 
that the spacetime geometry, and perhaps the spacetime points themselves too, 
should be introduced using the distance \emph{as the primary concept}. Two 
years later, Penrose initiated the project to derive the various geometric 
structures of spacetime from more fundamental quantum concepts. (The essay 
with his original ideas became available for the public much later, only in 
\cite{Pe79}.) 

As an illustration of his strategy, using combinatorial techniques, Penrose 
showed how the \emph{conformal structure} (i.e. the geometry of 
\emph{directions}) of the Euclidean 3-space can be recovered from the 
quantum mechanics of spins \cite{Penrose}. However, apart from the 
dissertation \cite{Mo}, this project was not continued and has not been 
completed. 

Using standard techniques of quantum mechanics, in our previous papers, we 
re-derived the conformal structure of the Euclidean 3-space \cite{Sz21c}, 
and derived the \emph{metric} structure of the Euclidean 3-space \cite{Sz22a} 
and of the Minkowski spacetime \cite{Sz23a} from $SU(2)$, Euclidean and 
Poincar\'e invariant quantum mechanical systems, respectively. In the 
present paper, we complete the first stage of the (then) `radically new' 
project of Penrose in the spirit of Wheeler's `radical conservatism' 
\cite{Th}, by determining \emph{formal} conditions under which the (local) 
geometry of general curved spacetimes with $C^2$ Lorentzian metrics can be 
recovered in the classical limit from abstract Poincar\'e-invariant quantum 
\emph{mechanics}. In particular, we use \emph{only} ideas and techniques of 
standard general relativity, differential geometry and Poincar\'e-invariant 
quantum mechanics. (Following the classical paper \cite{EPS}, by spacetime 
we mean only a four dimensional Lorentzian geometry, but still no field 
equation is used. Like in the foundation of the classical theory in 
\cite{EPS} [see also \cite{Wo,Per,Linn1}], \emph{mechanical} concepts 
appear to be enough to introduce the spacetime geometry. \emph{Field 
theoretical} concepts and ideas are expected to be needed only in the 
introduction of the field equations.) 

The present investigations are based in an essential way on our previous 
results \cite{Sz23a}, in which the metric structure of the Minkowski 
spacetime has been recovered in the classical limit from an expression 
called `empirical distance'. The latter is built from the basic quantum 
observables of Poincar\'e-invariant elementary quantum mechanical systems 
in an abstract, algebraic formulation of quantum mechanics (see e.g. 
\cite{Haag,FeRe,BuFre}) \emph{without assuming anything about space or 
spacetime}. (At this point it should be stressed that the Poincar\'e 
symmetry is the symmetry of the \emph{algebra of the basic observables} of 
the quantum mechanical systems, rather than that of the spacetime.) 

The key idea that we adopt here is the `localization' of the concepts of 
\cite{Sz23a}, just as how the flat pseudo-Euclidean spaces are `localized' 
and became the geometry of tangent spaces of general curved pseudo-Riemannian 
geometries. We have no a priori assumption on the `micro-structure' of the 
spacetime. We take general relativity as it is as the correct 
\emph{classical} theory of spacetime and gravity; and also the abstract, 
algebraic formulation of (Poincar\'e-invariant) quantum mechanics (without 
referring to any notion of spacetime) as the most successful theory of 
quantum \emph{particles}. (In the context of quantum mechanics, by 
Poincar\'e group $E(1,3)$ we mean the semidirect product of $SL(2,
\mathbb{C})$ and the group of the four dimensional translation group, 
rather than the isometry group of the Minkowski space.) In the present 
paper, we search only for the quantum states of the Poincar\'e-invariant 
elementary quantum mechanical systems by means of which the known 
\emph{local} geometric structures (in particular, the metric) of the 
classical spacetime can \emph{formally} be recovered from the observables 
of the quantum systems \emph{in their classical limit}. Here, by `local 
structures' we mean those in convex normal neighbourhoods (in particular 
the curvature tensor and all the structures, e.g. the metric, that it 
determines locally), and by `formal' that still we do not intend to 
identify the source where the resulting quantum states may come from. The 
search for the roots of the origin of these states in the quantum systems 
themselves is the subject of a separate project. 

The general strategy of these investigations, which at some point deviates 
from that of \cite{Pe79}, has been discussed and summarized in the 
introduction of \cite{Sz23a}. Thus we do not repeat it here, and we 
concentrate only on the new aspects of the strategy. In particular, the 
present localization program rests on two new elements: 

First, as is well known in differential geometry (see e.g. 
\cite{KN1,Hicks,Milnor,Spivak2,ONeill}), the curvature tensor at a given 
point is completely determined by the sectional curvatures there, and the 
latter are linked directly to lengths of triplets of certain curves, 
forming a `triangle' on 2-surfaces generated by geodesics emanated from the 
given point. (The differential geometric background of Regge's non-smooth 
formulation of the spacetime geometry \cite{Regge}, and also of the 
strategy how to synthesize the `empirical geometry' from distances between 
\emph{points} \cite{MW}, is provided by these ideas.) Thus, to be able to 
recover the curvature from the quantum observables, first we should 
reformulate these geometric ideas and results in terms of \emph{distances} 
(i.e. by length of \emph{geodesics}, rather than only by length of certain 
non-geodesic curves like in the standard differential geometric texts above) 
between spacetime \emph{points}. 

Second, in Minkowski space the empirical distance, introduced in \cite{Sz23a}, 
is a distance between \emph{timelike straight lines} (and are realized as 
the length of certain \emph{spacelike straight line segments}), rather than 
between points. However, to be able to recover e.g. the curvature at some 
\emph{point} of the classical spacetime manifold using the same notion of 
distance, the notion of points should also be reformulated in the form that 
could be implemented in quantum theory. A point of the classical spacetime 
manifold can be characterized by the set of the timelike geodesics that 
intersect one another at that point, and the characteristic property of 
these geodesics is that the (spatial) empirical distance between any two of 
them is zero. (The idea how to represent spacetime points in this way is 
borrowed from twistor theory \cite{PeMacC72,HT,PR}, where a point of the 
Minkowski space is represented by the \emph{null} straight lines that 
intersect one another at that point.) It is this equivalent property that 
can be imported into quantum theory to \emph{define} `points' of the 
resulting `empirical, physical spacetime', obtained in the classical limit: 
points will emerge as \emph{pairs of timelike geodesics with vanishing 
empirical distance between them}. 

Technically, the localization is based on the use of convex normal 
neighbourhoods, and that all the geometric structures there can be mapped 
diffeomorphically into the tangent space of a point of these neighbourhoods. 
(As we will see, the ambiguity in such a diffeomorphism can be controlled 
by a pair $(\Lambda^a{}_b,\xi^a)$ of a Lorentz transformation and a 
`translation', just like in the Cartesian coordinate systems in Minkowski 
spacetime, though the set of these pairs do not have any obvious group 
structure). Hence, by this diffeomorphism the \emph{curved} spacetime metric 
emerges as an extra structure on the \emph{flat}, Minkowskian tangent space, 
and this makes it possible to link it to observables of abstract 
Poincar\'e-invariant quantum mechanical systems. 

In the present paper, we show that for any `geodesic triangle' above we can 
always find Poincar\'e-invariant quantum mechanical systems and a sequence 
of their states, defining the classical limit, such that the empirical 
distance, evaluated in the tensor product of these states, reproduces the 
lengths of the geodesics of the geodesic triangle in the classical limit. 
Therefore, \emph{the local metric structure of any (as we will see, $C^2$) 
Lorentzian geometry can be recovered from Poincar\'e-invariant quantum 
mechanics}. In the given context, the role of the curvature tensor is to 
shift the location of the image of the spacetime points in the tangent 
space. A different interpretation and role of the curvature in quantum 
theory will be given elsewhere. 

In Section \ref{sec-2}, we review the geometry of convex normal 
neighbourhoods and derive the form of the result on the geodesic triangles 
that we need. In the first half of Section \ref{sec-3}, we recall the idea 
behind the empirical distance and the key result in Minkowski spacetime, 
and we show how the spacetime points can be represented by pairs of timelike 
straight lines. In the second half of this section the previous concepts are 
generalized to curved spacetimes. Section \ref{sec-4} is devoted to the 
quantum mechanical considerations, where the key result, Theorem 
\ref{th-4.2.1}, is presented. The paper concludes with Section \ref{sec-5} 
with a few remarks and an Appendix where a technical question is clarified. 

Our sign conventions are those of \cite{HT,PR}. In particular, the signature 
of the spacetime metric is $(+,-,-,-)$, and the curvature is defined 
according to $-R^\alpha{}_{\beta\gamma\delta}X^\beta Z^\gamma W^\delta$ $:=\nabla_Z
(\nabla_WX^\alpha)-\nabla_W(\nabla_ZX^\alpha)-\nabla_{[Z,W]} X^\alpha$, where 
$[Z,W]^\alpha$ is the Lie bracket of the vectors. We use the units in which 
$c=1$, but we keep $\hbar$.


\section{The geometry of convex normal neighbourhoods}
\label{sec-2}

After summarizing the necessary differential geometric background (mostly 
to fix the notations and the differential geometric notions) in subsection 
\ref{sub-2.1}, we show in subsection \ref{sub-2.2} how the curvature tensor 
at an arbitrary point is related to the lengths of certain \emph{spacelike 
geodesics} between pairs of spacelike separated points.

\subsection{Convex normal neighbourhoods}
\label{sub-2.1}

Let $M$ be the base manifold of a Lorentzian 4-geometry, $q\in M$ and $U
\subset M$ a convex normal neighbourhood of $q$; i.e. for some open 
neighbourhood $V$ of the zero vector in the tangent space $T_qM$ the 
exponential map $\exp_q:V\to U$ is a diffeomorphism, and any two points 
of $U$ can be joined by a unique geodesic segment which lays entirely in 
$U$ (see e.g. \cite{KN1,Hicks,Milnor,Spivak2}). We parameterize the points 
of the geodesic defined by the exponential map according to $\alpha(t):=
\exp_q(tX)$ with fixed $X^\alpha\in T_qM$, and if $X^\alpha$ is a unit vector, 
then the natural (affine) parameter $t$ is the arc length parameter along 
$\alpha$. 

The Riemannian normal coordinate system, based on the point $q\in M$ and, 
say, an \emph{orthonormal} basis $\{E^\alpha_a\}$ in $T_qM$, $a=0,...,3$, 
with timelike $E^\alpha_0$, is defined by the exponential map: the 
coordinates of the point $r\in U$ are defined to be $x^a=(x^0,...,x^3)\in
\mathbb{R}^4$ if $r=\exp_q(x^aE_a)$. Here, the Greek indices are referring 
to a general coordinate system on $U$ (which can also be considered to be 
abstract tensor indices), and the small Latin indices are concrete name 
indices referring to the orthonormal basis $\{E^\alpha_a\}$ at $q$. Hence, 
the Riemannian normal coordinates are not only coordinates in the sense of 
differential topology, but (actually) they are adapted to the metric 
structure of the spacetime geometry, analogously to the Cartesian 
coordinates in (pseudo-) Euclidean spaces. In the coordinates $\{x^a\}$, 
the geodesics through $q$ in $U$ are straight lines in $V$ through the 
origin of $T_qM$: $x^a=tX^a$ for some vector $X^\alpha=X^aE^\alpha_a$ with 
the components $X^a$ in the basis $\{E^\alpha_a\}$; and the origin of this 
coordinate system corresponds to $q$. If the components of the curvature 
tensor in this basis at $q$ are $R_{abcd}$, then, as is well known, 
\begin{eqnarray}
&{}&g_{ab}(x)=\eta_{ab}-\frac{t^2}{3}R_{acdb}X^cX^d+O(t^3),
  \label{eq:2.1.1a}\\
&{}&\Gamma^a_{bc}(x)=\frac{t}{3}\bigl(R^a{}_{bcd}+R^a{}_{cbd}\bigr)X^d+
  O(t^2), \label{eq:2.1.1b}\\
&{}&R^a{}_{bcd}(x)=R^a{}_{bcd}+O(t) \label{eq:2.1.1c}
\end{eqnarray}
hold, where $\eta_{ab}:={\rm diag}(1,-1,-1-1)$. 
Note that, by (\ref{eq:2.1.1a})-(\ref{eq:2.1.1c}), the curvature at $q$ 
provides a control on the derivatives of the metric on the whole $U$ only up 
to second order. To have control on the first $(r+2)$ derivatives, $r\geq0$, 
the derivatives of the curvature up to $r$th order would have to be specified 
at $q$. This more accurate (but technically considerably more complicated) 
case will not be considered in the present paper. By (\ref{eq:2.1.1a}) the 
metric on $U$ can also be considered to be a perturbed flat metric, where the 
perturbation grows quadratically in the leading order with the coordinates 
$x^a=tX^a$, and the factor of proportionality is just the curvature. The 
significance of the presence of the \emph{generic} curvature in 
(\ref{eq:2.1.1a}) is that it provides a guarantee that a \emph{generic}, 
rather than only some special spacetime geometry is considered. 

If $Z^\alpha$ is any vector at $q$, then it can be extended to the whole 
$U$ in a unique way by parallelly propagating it along the (`radial') 
geodesics $\alpha$: using (\ref{eq:2.1.1b}), for its components $Z^a(t)$ at 
the point $\alpha(t)$ in the coordinate basis $\{(\partial/\partial x^a)
^\alpha\}$ we obtain 
\begin{equation}
Z^a(t)=\Bigl(\delta^a_b+\frac{t^2}{6}R^a{}_{cdb}X^cX^d+O(t^3)\Bigr)Z^b, 
\label{eq:2.1.2a}
\end{equation}
where $Z^a$ and $X^a$ are the components of $Z^\alpha$ and $X^\alpha$, 
respectively, at $q=\alpha(0)$ in the basis $\{E^\alpha_a\}=\{(\partial/
\partial x^a)^\alpha\vert_q\}$. In particular, the basis $\{E^\alpha_a\}$ can 
be extended in a unique way from $q$ to the whole $U$ by parallel transport 
via (\ref{eq:2.1.2a}). (N.B.: In general, by (\ref{eq:2.1.1a}), the 
coordinate basis $\{(\partial/\partial x^a)^\alpha\}$ is \emph{not} 
orthonormal on $U-\{q\}$, and hence, in particular, not parallelly 
propagated along the radial geodesics.) The above formula can be inverted 
for $Z^a$: 
\begin{equation}
Z^a=\Bigl(\delta^a_b-\frac{t^2}{6}R^a{}_{cdb}X^cX^d+O(t^3)\Bigr)Z^b(t),
\label{eq:2.1.2}
\end{equation}
by means of which a vector at any point of $U$ can be transported into $T_qM$. 
Since parallel transport preserves the metric, the norm of $Z^\alpha(t)$ at 
$r$, i.e. at $x^a=tX^a$, with respect to the \emph{physical metric} $g
_{\alpha\beta}(x)$ there, is equal to the norm of $Z^a$, obtained from 
(\ref{eq:2.1.2}) at $q$, with respect to the \emph{flat} metric (as one can 
check it explicitly using (\ref{eq:2.1.1a}) and (\ref{eq:2.1.2}), too).  

The Riemannian normal coordinates $x^a$ on $U$ are uniquely determined by 
the origin point $q$ and the orthonormal basis $\{E^\alpha_a\}$ at $q$. 
Hence all the ambiguities in these coordinates are manifested in the Lorentz 
transformations taking one orthonormal basis to another one in $T_qM$; and 
in the choice of another origin point $\tilde q\in U$. Since the new origin 
point can always be written in a unique way as $\tilde q=\exp_q(\xi)$ for 
some vector $\xi^\alpha\in T_qM$, the ambiguities in the Riemannian normal 
coordinates on $U$ can be controlled by pairs $(\Lambda^a{}_b,\xi^a)$ of 
Lorentz transformations and vectors, which might be called `translations'. 
(Nevertheless, the set of these pairs does not seem to have any obvious 
group structure.) The action of the former is obvious: if the Lorentz 
transformation takes $E^\alpha_a$ into $\tilde E^\alpha_a:=E^\alpha_b\Lambda^b
{}_a$, then by $r=\exp_q(x^aE_a)=\exp_q(\tilde x^a\tilde E_a)$ it maps $x^a$ 
into $\tilde x^a=(\Lambda^{-1})^a{}_bx^b$, and $\Lambda^a{}_b$ is unrestricted. 
However, the action of the `translation' $\xi^\alpha$ is less trivial, and 
$\xi^\alpha$ must be `small enough'. To see this, let $\{\tilde E^\alpha_a\}$ 
denote the basis at $\tilde q:=\exp_q(\xi)$ obtained from $\{E^\alpha_a\}$ by 
parallel transport via (\ref{eq:2.1.2a}), and denote the coordinates of the 
point $r=\exp_q(x^aE_a)$ in the Riemannian normal coordinate system based on 
$(\tilde q,\{\tilde E^\alpha_a\})$ by $\tilde x^a$; i.e. (by definition) $r=
\exp_{\tilde q}(\tilde x^a\tilde E_a)$ also holds. Then for the 
\emph{components} of the position vectors $x^bE^\alpha_b,\xi^\alpha\in T_qM$ 
and $\tilde x^b\tilde E^\alpha_b\in T_{\tilde q}M$ in the coordinate basis 
$\{(\partial/\partial x^a)^\alpha\}$ we have that $x^bE^a_b+O(\vert x\vert^2) 
=\xi^a+\tilde x^b\tilde E^a_b$, and hence by (\ref{eq:2.1.2a}) that 
\begin{equation}
\tilde x^a=\Bigl(\delta^a_b-\frac{1}{6}R^a{}_{cdb}\xi^c\xi^d+O(\vert\xi\vert^3)
\Bigr)x^b-\xi^a+O(\vert x\vert^2). \label{eq:2.1.3}
\end{equation}
In Minkowski space the position vector of $r$ with respect to $\tilde q$ 
is just $-1$ times the position vector of $\tilde q$ with respect to $r$. 
By equation (\ref{eq:2.1.3}) the curvature destroys this anti-symmetry of 
the translation and the position vectors, $\xi^\alpha$ and $x^aE^\alpha_a$. In 
fact, for the sum of (\ref{eq:2.1.3}) and of the equation obtained from 
(\ref{eq:2.1.3}) by interchanging the role of $\xi^\alpha$ and $x^aE^\alpha_a$ 
we obtain that, in the leading order, $\tilde x^a+\tilde\xi^a=\frac{1}{6}
R^a{}_{bcd}(x^b-\xi^b)\xi^cx^d$. In the presence of curvature, the concept of 
the position vector and of the translation splits even in convex normal 
neighbourhoods. 

Thus, to summarize, the curved geometry of $U$ is mapped diffeomorphically 
into the flat $T_qM$. The vector $tX^a$ in the parameterization of the point 
$r=\exp_q(tX)$ may be considered to be the `position vector' of the point 
$r$ with respect to $q$, all the tensors at $r$ can be transported into $q$ 
via (\ref{eq:2.1.2}), and the resulting tensor can be interpreted as a tensor 
at $tX^a$ in $T_qM$. On the other hand, due to the curvature, the concept of 
translations deviates from that of the position vectors. The curvature 
emerges as an extra structure on $T_qM$. 


\subsection{Curvature from spatial geodesic lengths}
\label{sub-2.2}

Let $X^\alpha,Y^\alpha\in T_qM$ be two unit \emph{spacelike} vectors that are 
orthogonal to each other, let $X^\alpha(w):=\cos wX^\alpha+\sin wY^\alpha$, 
and let us form the 2-surface $S$ in $U$ through $q$ whose points are of 
the form $\exp_q(tX(w))$. Thus, $S$ is formed by the 1-parameter family 
$\alpha_w$ of geodesics through $q$, defined by $\alpha_w(t):=\exp_q(tX(w))$ 
with $w$ as the family parameter. These geodesics are orthogonal to the 
1-parameter family of curves $\lambda_t$, defined by $\lambda_t(w):=\exp_q
(tX(w))$ with $t$ as the family parameter (`Gauss lemma' 
\cite{KN1,Hicks,Milnor,Spivak2}). In Riemann's normal coordinates, these 
curves are given by the arcs $\lambda^a_t(w)=tX^a(w)$. If the 2-surface $S$ 
were intrinsically flat, then the length of $\alpha_w$ between $q=\alpha
_w(0)$ and $\alpha_w(t)$ would be $t$ for any $w$, and the length of the arc 
$\lambda_t$ between $\alpha_0(t)$ and $\alpha_w(t)$ would be $wt$. In general, 
although the length of $\alpha_w$ is still $t$, but, as it is noted in an 
informal remark in \cite{Milnor}, page 101, the length of $\lambda_t$ is 
affected by $t^2/6$-times the component $R_{abcd}X^aY^bX^cY^d$ of the 
curvature tensor at $q$. The significance of this result is that, in 
Riemannian geometry, the given component of the curvature can be determined 
\emph{directly} from the length of $\alpha_w$ and $\lambda_t$. Since all 
the components of the curvature tensor is determined by its components of 
the form $R_{abcd}X^aY^bX^cY^d$, called the sectional curvature determined by 
$X^\alpha$ and $Y^\alpha$, the whole curvature at $q$ can be determined in 
this way by considering various 2-planes spanned by vectors like $X^\alpha$ 
and $Y^\alpha$ in $T_qM$. (For a more detailed discussion of these and related 
ideas in Riemannian geometry, see e.g. \cite{KN1,Hicks,Milnor,Spivak2}, and 
in semi-Riemannian geometries \cite{ONeill}.) 

Our aim is to find the relation between the curvature tensor of Lorentzian 
4-geometries in terms of lengths of \emph{spacelike geodesics}. However, the 
curves $\lambda_t$ are \emph{not} geodesics, and the question is whether or 
not an analogous result, involving the radial geodesics $\alpha_0$ and 
$\alpha_w$ and the uniquely determined \emph{geodesic} $\chi$ between the 
points $\alpha_0(t)=\lambda_t(0)$ and $\alpha_w(t)=\lambda_t(w)$, can be 
derived. Moreover, one should check that it is enough to use only 
\emph{spacelike} 2-planes to determine the whole curvature, even though the 
geometry is Lorentzian. The answer to the first question is given by the 
lemma below, and the second question is clarified in the Appendix. 

If the 2-surface $S$ were intrinsically flat, then the geodesic from 
$\alpha_0(t)$ to $\alpha_w(t)$ would be the uniquely determined straight 
line segment in $S$, i.e. the position vector pointing from $\alpha_0(t)$ 
to $\alpha_w(t)$. This position vector is $t(X^a(w)-X^a)=2t\sin(w/2)V^a$, 
where, as elementary calculation shows, its \emph{unit} tangent $V^a$ is 
the linear combination $V^a=-\sin(w/2)X^a+\cos(w/2)Y^a$. For sufficiently 
small $w$ the length of this position vector is $tw(1-w^2/4!+O(w^4))$. The 
effect of the curvature on this length is given by the next lemma, 
motivated by the remark in \cite{Milnor}, page 101, quoted above. 
\begin{lemma}\label{lemma-2.2.1}
Let $X^\alpha,Y^\alpha\in T_qM$ be unit spacelike vectors orthogonal to each 
other, and, for $t>0$ and $w\geq0$, let us form the geodesics $\alpha_0$, 
$\alpha_w$ and the curve $\lambda_t$. Let $\chi$ be the geodesic segment 
between the points $\alpha_0(t)=\lambda_t(0)$ and $\alpha_w(t)=\lambda_t(w)$. 
Then, for sufficiently small $t$ and $w$, the length of $\chi$ is 
\begin{equation}
L[\chi]=tw\Bigl(1-\frac{t^2}{6}R_{abcd}X^aY^bX^cY^d+O(t^3)+O(w)\Bigr).
\label{eq:2.2.1}
\end{equation}
\end{lemma}
\begin{proof}
Since $U$ is a convex normal neighbourhood of $q$, for sufficiently small 
$t$ and $w$ the geodesic $\chi$ exists, unique and lays in $U$. Let the 
curve $\lambda_t$ and the geodesic $\chi$ be given in the Riemannian normal 
coordinates by the functions $\lambda^a_t(\bar w)$ and $\chi^a(\bar w)$, 
respectively, where $0\leq\bar w\leq w$. Since the order of 
differentiability of the two terms in the geodesic equation, $({\rm d}^2
y^a/{\rm d}v^2)+\Gamma^a_{bc}(y(v))({\rm d}y^b/{\rm d}v)({\rm d}y^c/{\rm d}v)
=0$, must be the same, for a $C^k$ connection the differentiability class of 
the solution $y^a(v)$ must be $C^{k+2}$. Since by (\ref{eq:2.1.1b}) $\Gamma^a
_{bc}$ is $C^1$ in the coordinates $x^a$, $\lambda^a_t(\bar w)$ and $\chi^a
(\bar w)$ are $C^3$. Clearly, $\chi^a(0)=\lambda^a_t(0)$, and if $t\to0$, 
then $\chi^a(\bar w)-\lambda^a_t(\bar w)\to0$. Hence there are $C^2$ 
functions $w^a$ of $\bar w$ and $t$ such that $\chi^a(\bar w)-\lambda^a_t
(\bar w)=\bar w\,t\,w^a(\bar w,t)$ holds. Moreover, since in the $w\to0$ 
limit the tangent of $\chi$ at its starting point $\lambda_t(0)$ tends to 
the tangent of $\lambda_t$ there, there exist $C^2$ functions $W^a$ of 
$\bar w$ and $t$ such that 
\begin{equation*}
\frac{{\rm d}\chi^a(\bar w)}{{\rm d}\bar w}-\frac{{\rm d}\lambda^a_t
(\bar w)}{{\rm d}\bar w}=w\,t\,W^a(\bar w,t). 
\end{equation*}
Since the geodesic $\chi$ depends on $w$, the functions $w^a$ and $W^a$ 
depend on $w$, too. Using these formulae for $\chi^a(\bar w)$ and its 
$\bar w$-derivative and the fact that $\lambda^a_t(\bar w)=tX^a(\bar w)=t
(\cos\bar wX^a+\sin\bar wY^a)$, by (\ref{eq:2.1.1a}) we obtain 
\begin{eqnarray*}
g_{ab}(\chi)\!\!\!\!\!&{}\!\!\!\!\!&\frac{{\rm d}\chi^a}{{\rm d}\bar w}
  \frac{{\rm d}\chi^b}{{\rm d}\bar w}=\eta_{ab}\Bigl(\frac{{\rm d}
  \lambda^a_t}{{\rm d}\bar w}+wtW^a\Bigr)\Bigl(\frac{{\rm d}\lambda^b_t}
  {{\rm d}\bar w}+wtW^b\Bigr)+ \\
+\!\!\!\!\!&{}\!\!\!\!\!&\frac{1}{3}R_{abcd}\Bigl(\frac{{\rm d}\lambda^a_t}
  {{\rm d}\bar w}+wtW^a\Bigr)\Bigl(\lambda^b_t+\bar wtw^b\Bigr)\Bigl(
  \frac{{\rm d}\lambda^c_t}{{\rm d}\bar w}+wtW^c\Bigr)\Bigl(\lambda^d_t+
  \bar wtw^d\Bigr)+O\Bigl(\vert\frac{{\rm d}\chi^e}{{\rm d}\bar w}\vert^2
  \Bigr)O(t^3)=\\
=t^2\Bigl\{\!\!\!\!\!&{}\!\!\!\!\!&-1+\frac{t^2}{3}R_{abcd}X^aY^bX^cY^d+
  wF_1+w^2F_2+ \\
+\!\!\!\!\!&{}\!\!\!\!\!&t^2\Bigl(wG_1+\bar wG_2+w^2g_3+w\bar wG_4+\bar w^2
  G_5+w\bar w^2G_6+w^2\bar wG_7+w^2\bar w^2G_8\Bigr)+O(t^3)\Bigr\}, 
\end{eqnarray*}
where $F_1$, $F_2$, $G_1$, ..., $G_8$ are $C^2$ functions of $w$, $\bar w$ 
and $t$. The Taylor expansion of the square root of its absolute value with 
respect to $t$, $w$ and $\bar w$ around $t=w=\bar w=0$ up to second order is 
\begin{equation*}
\sqrt{\vert g_{ab}(\chi)\frac{{\rm d}\chi^a}{{\rm d}\bar w}\frac{{\rm d}\chi^b}
{{\rm d}\bar w}\vert}=t\Bigl(1-\frac{t^2}{6}R_{abcd}X^aY^bX^cY^d+O(w)+O(\bar w^2)
+O(t^3)\Bigr),
\end{equation*}
and hence its integral with respect to $\bar w$ from $0$ to $w$ gives 
(\ref{eq:2.2.1}). 
\end{proof}

Thus, the component $R_{abcd}X^aY^bX^cY^d$ of the curvature tensor at $q$ is 
determined by the length of $\chi$ in the triplet $(\alpha_0,\alpha_w,\chi)$ 
of \emph{spacelike geodesics}; and, by (\ref{eq:2.2.1}), it is given by the 
derivatives 
\begin{equation}
R_{abcd}X^aY^bX^cY^d=-\Bigl(\frac{\partial^4}{\partial t^3\partial w}
L[\chi]\Bigr)_{t=w=0}=-\frac{1}{16}\Bigl(\frac{\partial^6}{\partial t^4
\partial w^2}(L[\chi])^2\Bigr)_{t=w=0}. \label{eq:2.2.2}
\end{equation}
In the appendix we show that \emph{all} the components of the curvature 
tensor are determined by the contractions of the form $R_{abcd}X^aY^bX^cY^d$ 
for 21 (in fact, only 20 independent) appropriately chosen \emph{purely 
spacelike} pairs $(X^\alpha,Y^\alpha)$. 


\section{Lorentzian 4-geometries from classical empirical distances}
\label{sec-3}

In subsection \ref{sub-3.1} we summarize the key ideas behind the empirical 
distance, and it will be shown how the concept of spacetime points can be 
characterized, or rather to be defined operationally, using this distance. 
In subsection \ref{sub-3.2}, we consider distances between \emph{points} of 
convex normal neighbourhoods in a \emph{curved} spacetime, and then we 
reinterpret these distances as distances between certain \emph{timelike 
geodesics}. This reinterpretation of distances yields that the spacetime 
points can be introduced \emph{operationally} as pairs of timelike geodesics 
with zero distance between them; and it is this form of the distance between 
points that can be carried over into quantum theory.

\subsection{The distance between timelike straight lines in Minkowski 
space}
\label{sub-3.1}

\subsubsection{The empirical distance}
\label{sub-3.1.1}

In Minkowski space the distance between any two non-parallel timelike 
straight lines is well defined. In fact, if the two straight lines are 
(e.g.) future directed, then, in \emph{Cartesian coordinates} $x^a=
(x^0,...,x^3)$, they can be given by $\gamma^a_1(u)=\Lambda^a_1{}_b\gamma^b_0
(u)+\xi^a_1$ and $\gamma^a_2(u)=\Lambda^a_2{}_b\gamma^b_0(u)+\xi^a_2$, 
respectively, for some Lorentz boosts $\Lambda^a_1{}_b$ and $\Lambda^a_2{}_b$ 
and translations $\xi^a_1$ and $\xi^a_2$, where $\gamma^a_0(u):=u\delta^a_0$ 
and $u$ is the proper time parameter. Then the (square of the) distance 
$D(\gamma_1,\gamma_2)$, or simply $D_{12}$, between these two straight lines 
is 
\begin{equation}
(D_{12})^2:=-\Pi_{ab}\bigl(\xi^a_1-\xi^a_2\bigr)\bigl(\xi^b_1-\xi^b_2\bigr),
\label{eq:3.1.1}
\end{equation}
where (considering these straight lines to be the world lines of classical 
freely moving point particles with positive rest mass $\mu_1$ and $\mu_2$, 
respectively, and hence their energy-momentum 4-vectors are $p^a_1=\mu_1
\Lambda^a_1{}_0$ and $p^a_2=\mu_2\Lambda^a_2{}_0$) $\Pi^a_b$ is the projection 
to the spacelike 2-plane orthogonal to $p^a_1$ and $p^a_2$ given explicitly 
by 
\begin{eqnarray}
\Pi^a_b:=\!\!\!\!\!&{}\!\!\!\!\!&\delta^a_b+\frac{1}{P^4_{12}-\mu^2_1\mu^2_2}
  \Bigl(\mu^2_2p^a_1p_{1b}+\mu^2_1p^a_2p_{2b}-P^2_{12}(p^a_1p_{2b}+p^a_2p_{1b})
  \Bigr)= \label{eq:3.1.2} \\
=\!\!\!\!\!&{}\!\!\!\!\!&-\frac{1}{P^4_{12}-\mu^2_1\mu^2_2}\bigl(\varepsilon
  ^a{}_{ecd}p^c_1p^d_2\bigr)\bigl(\varepsilon^e{}_{bgh}p^g_1p^h_2\bigr)
  =\frac{1}{((\Lambda^{-1}_1\Lambda_2)_{00})^2-1}(\varepsilon^a{}_{cde}\Lambda
  ^c_1{}_0\Lambda^d_2{}_0)(\varepsilon_{bgh}{}^e\Lambda^g_1{}_0\Lambda^h_2{}_0). 
  \nonumber
\end{eqnarray}
Here $\varepsilon_{abcd}$ is the volume 4-form (which, actually, is the 
totally skew Levi-Civita symbol), and $P^2_{12}:=\eta_{ab}p^a_1p^b_2$. 
This $D_{12}$ is just the well defined Lorentzian length of the straight 
line segment in this spacelike 2-plane between a uniquely determined point 
$\nu_1$ of $\gamma_1$ and a uniquely determined point $\nu_2$ of $\gamma_2$. 
(If the straight lines $\gamma_1$ and $\gamma_2$ were parallel, then the 
denominator in (\ref{eq:3.1.2}) would be zero. However, by taking limits, 
the distance $D_{12}$ could be extended in a unique [and obvious] way to 
pairs of parallel timelike straight lines, too. In this case, the points 
$\nu_1$ and $\nu_2$ would not be uniquely determined.) Here $\Pi^a_b(\xi^b_2-
\xi^b_1)$ can be interpreted as the relative position vector pointing from 
$\nu_1$ to $\nu_2$, or the translation taking $\nu_1$ into $\nu_2$. The 
significance of this distance $D_{12}$ is twofold. 

The first is that (the square of) this distance can be re-expressed in 
terms of the basic observables of Poincar\'e-invariant elementary classical 
mechanical systems, viz. the 4-momenta $p^a_1$ and $p^a_2$ and the angular 
momenta $J^{ab}_1$ and $J^{ab}_2$, and the structure of this alternative 
expression, denoted by $d^2_{12}$, showed how to construct the analogous 
expression \emph{in quantum theory}. In the classical theory, the 
construction of $d^2_{12}$ is as follows. 

If the rest mass $\mu$ of such a system is positive, then one can always 
find a 1-parameter family of translations, viz. $\xi^a=-M^a/\mu^2+up^a/\mu$, 
$u\in\mathbb{R}$, by means of which the centre-of-mass vector, $M^a:=J^{ab}
p_b$, can be taken to be vanishing. (Recall that under the translation with 
$\xi^a$ the basic observables transform as $(p^a,J^{ab})\mapsto(p^a,J^{ab}+
\xi^ap^b-\xi^bp^a)$; and hence, under such a translation, $M_a\mapsto
\tilde M_a:=M_a+(\mu^2\eta_{ab}-p_ap_b)\xi^b$. N.B.: This translation is $-1$ 
times the translation in the parameterization of the timelike straight lines 
given in the beginning of this subsection, and hence the latter is $M^a/
\mu^2$ up to the addition of an arbitrary term of the form $up^a/\mu$.) 
These translations point to the points of a timelike straight line in 
Minkowski space, the so-called centre-of-mass world line of the elementary 
mechanical system. Then, considering the straight lines $\gamma_1$ and 
$\gamma_2$ to be the centre-of-mass world lines of two Poincar\'e-invariant 
elementary classical mechanical systems, (\ref{eq:3.1.1}) with these special 
translations yields an alternative form of this distance in terms of the 
4-momentum and the angular momentum of the elementary systems: this is the 
square $(d_{12})^2:=-\eta_{ab}d^a_{12}d^b_{12}$ of the relative position vector 
\begin{equation}
d^a_{12}:=\Pi^a_b\Bigl(\frac{M^b_1}{\mu^2_1}-\frac{M^b_2}{\mu^2_2}\Bigr)=
-\frac{1}{P^4_{12}-\mu^2_1\mu^2_2}\varepsilon^a{}_{bcd}p^b_1p^c_2\Bigl(S^d_{12}
-P^2_{12}\bigl(\frac{S^d_1}{\mu^2_1}-\frac{S^d_2}{\mu^2_2}\bigr)\Bigr), 
\label{eq:3.1.3}
\end{equation}
where $S^a:=\frac{1}{2}\varepsilon^a{}_{bcd}J^{bc}p^d$ is the Pauli--Lubanski 
spin vector and $S^a_{12}:=\frac{1}{2}\varepsilon^a{}_{bcd}(J^{bc}_1p^d_2+J^{bc}
_2p^d_1)$. (For the details, see \cite{Sz23a}.) Thus, the distance between 
any two timelike straight lines has been re-expressed by \emph{observables 
of classical mechanical systems}. 

As we noted above, the significance of $d_{12}$ (that we call the `empirical 
distance') is that its structure showed how to construct an analogous 
empirical distance from the \emph{basic quantum mechanical observables} by 
means of which the metric structure of the Minkowski space could be recovered 
in the classical limit. In fact, denoting this quantum mechanical expression 
(which is evaluated in the quantum state of the systems) also by $d_{12}$, 
the following theorem \cite{Sz23a} could be proven: 
\begin{theorem}\label{th-3.1.1}
Let $\gamma_{\bi}$, ${\bi}=1,...,N$, be timelike straight lines in Minkowski 
space such that no two of them are parallel. Then there are $N$ 
Poincar\'e-invariant quantum mechanical systems and a sequence $\phi_{{\bi}k}$, 
$k\in\mathbb{N}$, of pure quantum states of them such that, in the $k\to
\infty$ limit, the empirical distances $d_{\bi\bj}$, calculated in the pure 
tensor product states $\phi_{1k}\otimes\cdots\otimes\phi_{Nk}$, tend with 
asymptotically vanishing uncertainty to the classical Lorentzian distances 
$D_{\bi\bj}$ between $\gamma_{\bi}$ and $\gamma_{\bj}$ for any ${\bi,\bj}=
1,...,N$. 
\end{theorem}
\noindent
(A more detailed discussion of the concepts here will be given in subsection 
\ref{sub-4.1}.) The primarily aim of the present paper is to find an 
analogous result on how the $C^2$ Lorentzian metric on a neighbourhood of a 
point of the classical spacetime can be recovered. 


\subsubsection{Points of Minkowski spacetime as derived concepts}
\label{sub-3.1.2}

The other significance of $D_{12}$ (and hence of $d_{12}$) is that this makes 
it possible \emph{to determine the spacetime points} in an \emph{operational 
way}. (This issue was not considered in \cite{Sz23a}.) Theorem \ref{th-3.1.1} 
is about how the distance between (timelike) \emph{straight lines}, rather 
than between points, can be recovered from quantum mechanics. Thus, if we 
want to recover distances between \emph{points}, the notion of point must 
also be reformulated in a way to be able to implement it in the quantum 
theory. 

Clearly, any point $q$ of Minkowski space is specified completely by the set 
$[q]$ of (e.g. the future directed) timelike straight lines through $q$. 
Physically, this means that the event represented by the point $q$ is 
identified to be just the meeting of the freely moving massive particles 
with the timelike straight lines as their world lines. But $D_{12}=0$ holds 
precisely when $\gamma_1$ and $\gamma_2$ intersect each other at some 
(uniquely determined) point. Hence the points of the Minkowski spacetime are 
in a one-to-one correspondence with the pairs of timelike straight lines 
with linearly independent tangents and vanishing distance between them. 
Moreover, the distance $D$ provides a criterion when a timelike straight 
line $\gamma$ passes through the intersection point of $\gamma_1$ and 
$\gamma_2$, and hence when two spacetime points, defined by pairs of 
timelike straight lines, coincide. 

Defining the set $[\gamma]:=\{\gamma'\,\vert \, D(\gamma,\gamma')=0\,\}$, 
i.e. the set of the timelike straight lines $\gamma'$ that intersect 
$\gamma$ somewhere, this criterion is given by the following proposition 
and its corollary. 
\begin{proposition}\label{proposition-3.1.1}
Let $\gamma_1$, $\gamma_2$ and $\gamma_3$ be timelike straight lines with 
\emph{linearly independent} tangents. Then $\gamma_1$, $\gamma_2$ and 
$\gamma_3$ intersect one another at a single point, say $q$, precisely 
when $D(\gamma_1,\gamma_2)=D(\gamma_2,\gamma_3)=D(\gamma_3,\gamma_1)=0$ 
holds. The set $[q]$ of the timelike straight lines through this common 
intersection point is $[\gamma_1]\cap[\gamma_2]\cap[\gamma_3]$. 
\end{proposition}
\begin{proof}
Clearly, if $\gamma_1$, $\gamma_2$ and $\gamma_3$ intersect one another at 
a single point, then $D(\gamma_1,\gamma_2)=D(\gamma_2,\gamma_3)=D(\gamma_3,
\gamma_1)=0$ holds. Conversely, let us suppose that $D(\gamma_1,\gamma_2)=0$. 
Then there is a uniquely determined point $q:=\gamma_1\cap\gamma_2$, and the 
two straight lines $\gamma_1$ and $\gamma_2$ lay in a timelike 2-plane. 
Since the tangent of $\gamma_3$ is linearly independent of that of $\gamma_1$ 
and of $\gamma_2$, the straight line $\gamma_3$ cannot lay in the same 
2-plane. Hence, $\gamma_3$ can intersect this timelike 2-plane in a single 
point. But by $D(\gamma_1,\gamma_3)=D(\gamma_2,\gamma_3)=0$ the straight 
line $\gamma_3$ must intersect both $\gamma_1$ and $\gamma_2$, which can 
happen only if its intersection point with the timelike 2-plane is just the 
intersection point $q$ of $\gamma_1$ and $\gamma_2$. Thus $q=\gamma_1\cap
\gamma_2\cap\gamma_3$. 

If $\gamma$ is any straight line through $q$, then by the definition of $q$ 
the line $\gamma$ intersects all of $\gamma_1$, $\gamma_2$ and $\gamma_3$, 
i.e. $\gamma\in[\gamma_1]\cap[\gamma_2]\cap[\gamma_3]$, and hence $[q]\subset
[\gamma_1]\cap[\gamma_2]\cap[\gamma_3]$. Conversely, let $\gamma\in[\gamma_1]
\cap[\gamma_2]\cap[\gamma_3]$. Then since $\gamma_1$, $\gamma_2$ and $\gamma
_3$ do not lay in the same 2-plane, $\gamma$ can intersect all of $\gamma_1$, 
$\gamma_2$ and $\gamma_3$ only if it intersects them at $q$, i.e. $[\gamma_1]
\cap[\gamma_2]\cap[\gamma_3]\subset[q]$. 
\end{proof}
\begin{corollary}\label{cor-3.1.1}
Let $\gamma_1$, $\gamma_2$ be timelike straight lines that cross each other 
at $q$, let $\gamma'_1$, $\gamma'_2$ be timelike straight lines that cross 
each other at $q'$, and suppose that at least three of the tangent of these 
straight lines are linearly independent. Then $q=q'$ if and only if 
$D(\gamma_1,\gamma'_1)=D(\gamma_1,\gamma'_2)=D(\gamma_2,\gamma'_1)=
D(\gamma_2,\gamma'_2)=0$. 
\end{corollary}
\begin{proof}
If $q=q'$, then the distance between any two of the timelike straight lines 
is clearly zero. 

Conversely, suppose that $D(\gamma_1,\gamma'_1)=D(\gamma_1,\gamma'_2)=
D(\gamma_2,\gamma'_1)=D(\gamma_2,\gamma'_2)=0$ holds, and that e.g. the 
tangent of $\gamma_1$, $\gamma_2$ and $\gamma'_1$ are linearly independent. 
Then by Proposition \ref{proposition-3.1.1} the straight line $\gamma'_1$ 
intersects $\gamma_1$ and $\gamma_2$ precisely at $q$. But $\gamma'_1$ does 
not lay in the 2-plane of $\gamma_1$ and $\gamma_2$, and hence $\gamma'_2$ 
can intersect the other three straight lines only at $q$. Hence $q=q'$. 
\end{proof}
By this corollary $D(\gamma_1,\gamma'_1)=D(\gamma_1,\gamma'_2)=D(\gamma_2,
\gamma'_1)=D(\gamma_2,\gamma'_2)=0$ and $q\not=q'$ can happen only in 
rather exceptional cases, viz. when all these straight lines lay in a single 
timelike 2-plane, and hence it gives a criterion when two points, specified 
by two pairs of timelike straight lines, coincide. Therefore, it provides 
an \emph{operational definition} of the spacetime points in terms of the 
distance $D$ and the 4-momenta. This representation of points is analogous 
to the representation of points by null geodesics and the orthogonality of 
their null twistor representatives in twistor theory \cite{PeMacC72,HT}. 

Using this representation of the points of the Minkowski space, the 
distance between any three \emph{spacelike} separated points, say $q$, $r_1$ 
and $r_2$, can be rewritten as the empirical distance between certain 
timelike straight lines through these points as follows. Let the points $q$, 
$r_1$ and $r_2$ be represented by the position vectors $\xi^a_{01}$, $\xi^a
_{11}$ and $\xi^a_{21}$, respectively, and hence $\xi^a_{11}-\xi^a_{01}$, 
$\xi^a_{21}-\xi^a_{01}$ and $\xi^a_{21}-\xi^a_{11}$ are spacelike. Let $\gamma^a
_{01}$ and $\gamma^a_{02}$ be different timelike straight lines through $q$, 
given in the Cartesian coordinate system, and, analogously, $\gamma^a_{11}$ 
and $\gamma^a_{12}$ through $r_1$, and $\gamma^a_{21}$ and $\gamma^a_{22}$ 
through $r_2$. Thus the points $q$, $r_1$ and $r_2$ are represented by the 
pairs $(\gamma^a_{01},\gamma^a_{02})$, $(\gamma^a_{11},\gamma^a_{12})$ and 
$(\gamma^a_{21},\gamma^a_{22})$, respectively. Clearly, these straight lines 
can be obtained from the single timelike straight line $\gamma^a_0(u)=u\,
\delta^a_0$ by an appropriate Poincar\'e transformation. In particular, 
$\gamma^a_{01}(u)=\Lambda^a_{01}{}_b\gamma^b_0(u)+\xi^a_{01}$ and $\gamma^a
_{02}(u)=\Lambda^a_{02}{}_b\gamma^b_0(u)+\xi^a_{01}$ hold for some Lorentz 
boosts $\Lambda^a_{01}{}_b$ and $\Lambda^a_{02}{}_b$; and we have analogous 
expressions for $\gamma^a_{11}(u)$, $\gamma^a_{12}(u)$, $\gamma^a_{21}(u)$ 
and $\gamma^a_{22}(u)$, too. 

To recover the distance between e.g. the points $q$ and $r_1$ as the 
empirical distance between e.g. $\gamma^a_{01}$ and $\gamma^a_{11}$ according 
to subsection \ref{sub-3.1.1}, these two straight lines should not be 
parallel and both must be orthogonal to $\xi^a_{11}-\xi^a_{01}$. Moreover, if 
this distance is expected to be the same between any of the two straight 
lines chosen one from the pair $(\gamma^a_{01},\gamma^a_{02})$ and the other 
from $(\gamma^a_{11},\gamma^a_{12})$, then no two of them may be parallel and 
any of them must be orthogonal to $\xi^a_{11}-\xi^a_{01}$. In terms of the 
Lorentz boosts and the relative position vector $\xi^a_{11}-\xi^a_{01}$ the 
latter conditions take the form 
\begin{equation}
\bigl(\xi^a_{11}-\xi^a_{01}\bigr)\eta_{ab}\Lambda^b_{01}{}_0=\bigl(\xi^a_{11}-
\xi^a_{01}\bigr)\eta_{ab}\Lambda^b_{02}{}_0=\bigl(\xi^a_{11}-\xi^a_{01}\bigr)
\eta_{ab}\Lambda^b_{11}{}_0=\bigl(\xi^a_{11}-\xi^a_{01}\bigr)\eta_{ab}\Lambda^b
_{12}{}_0=0. \label{eq:3.1.5}
\end{equation}
To recover the distance between $q$ and $r_2$, and also between $r_1$ and 
$r_2$, we obtain analogous conditions for $\xi^a_{21}-\xi^a_{01}$ and the 
boosts $\Lambda^a_{01}{}_0$, $\Lambda^a_{02}{}_0$, $\Lambda^a_{21}{}_0$, 
$\Lambda^a_{22}{}_0$; and for $\xi^a_{21}-\xi^a_{11}$ and the boosts $\Lambda^a
_{11}{}_0$, $\Lambda^a_{12}{}_0$, $\Lambda^a_{21}{}_0$, $\Lambda^a_{22}{}_0$. 
Since actually (because of the absolute parallelism in Minkowski space) the 
spacelike 2-plane spanned by $\xi^a_{11}-\xi^a_{01}$ and $\xi^a_{21}-\xi^a_{01}$ 
at $q$, by $\xi^a_{11}-\xi^a_{01}$ and $\xi^a_{21}-\xi^a_{11}$ at $r_1$, and by 
$\xi^a_{21}-\xi^a_{01}$ and $\xi^a_{21}-\xi^a_{11}$ at $r_2$ coincide, these 
orthogonality conditions are just the requirement that the boosts $\Lambda^a
_{01}{}_0$, $\Lambda^a_{02}{}_0$, $\Lambda^a_{11}{}_0$, $\Lambda^a_{12}{}_0$, 
$\Lambda^a_{21}{}_0$ and $\Lambda^a_{22}{}_0$, as vectors, be orthogonal to 
this spacelike 2-plane. Since the dimension of the spacetime is four, and 
hence the orthogonal complement of this 2-plane is a timelike 2-plane, all 
these boosts can be chosen to be different, and hence all the algebraic 
conditions can be imposed. It is this geometric picture that we carry over 
into the general, curved spacetime, and then also into the quantum theory.


\subsection{The distance between timelike geodesics in curved spacetimes}
\label{sub-3.2}

\subsubsection{Distances as Lorentzian norms}
\label{sub-3.2.1}

Although in Lorentzian geometries the notion of distance between spacelike 
separated points is not well defined in general, the distance between 
spatially separated points \emph{in convex normal neighbourhoods} can be 
introduced as the length of the \emph{uniquely determined} spacelike 
geodesic segment between them. In particular, with the notations and the 
result of Lemma \ref{lemma-2.2.1}, the square of the distance between the 
points $q$, $r_1:=\exp_q(tX)$ and $r_2:=\exp_q(tX(w))$ is given by 
\begin{eqnarray}
(D_{q,r_1})^2\!\!\!\!\!&=\!\!\!\!\!&t^2=-\eta_{ab}\bigl(tX^a\bigr)\bigl(tX^b
  \bigr), \label{eq:3.2.1a} \\
(D_{q,r_2})^2\!\!\!\!\!&=\!\!\!\!\!&t^2=-\eta_{ab}\bigl(tX^a(w)\bigr)\bigl(t
  X^b(w)\bigr), \label{eq:3.2.1b} \\
(D_{r_1,r_2})^2\!\!\!\!\!&=\!\!\!\!\!&t^2w^2\Bigl(1-\frac{t^2}{3}R_{abcd}
  X^aY^bX^cY^d+O(t^3)+O(w)\Bigr). \label{eq:3.2.1c}
\end{eqnarray}
Thus, the distances $D_{q,r_1}$ and $D_{q,r_2}$ are just the norms of the 
`position vectors' $tX^a$ and $tX^a(w)$, respectively, with respect to the 
\emph{flat} Minkowski metric $\eta_{ab}$. We show that, apart from higher 
order corrections, all these distances can be written in the same form. 

As we saw in subsection \ref{sub-2.2}, in Minkowski space the position 
vector of $r_2$ with respect to $r_1$ is $t(X^a(w)-X^a)$, where $X^a(w):=
\cos wX^a+\sin wY^a$. Considering this vector to be $Z^a(t)$ in 
(\ref{eq:2.1.2}), its parallel transport from $r_1$ to $q$ along the 
geodesic $\exp_q(tX)$ is given by 
\begin{eqnarray}
Z^a\!\!\!\!\!&=\!\!\!\!\!& t\Bigl((\cos w-1)X^a+\sin w(Y^a-\frac{t^2}{6}
  R^a{}_{bcd}X^bX^cY^d)+O(w)O(t^3)\Bigr)= \nonumber \\ 
\!\!\!\!\!&=\!\!\!\!\!&tw\Bigl(Y^a-\frac{t^2}{6}R^a{}_{bcd}X^bX^cY^d-\frac{w}
  {2}X^a+O(t^3)+O(w^2)\Bigr). \label{eq:3.2.2}
\end{eqnarray}
This motivates to consider the position vectors of the form 
\begin{eqnarray}
\hskip -35pt&{}&\xi^a_1:=tX^d(-\frac{w}{2})\Bigl(\delta^a_d-\frac{t^2}{6}
  R^a{}_{bcd}X^bX^c\Bigr)=t\Bigl(\cos\frac{w}{2}X^a-\sin\frac{w}{2}
  \bigl(Y^a-\frac{t^2}{6}R^a{}_{bcd}X^bX^cY^d\bigr)\Bigr), \label{eq:3.2.3a} \\ 
\hskip -35pt&{}&\xi^a_2:=tX^d(\frac{w}{2})\Bigl(\delta^a_d-\frac{t^2}{6}
  R^a{}_{bcd}X^bX^c\Bigr)=t\Bigl(\cos\frac{w}{2}X^a+\sin\frac{w}{2}
  \bigl(Y^a-\frac{t^2}{6}R^a{}_{bcd}X^bX^cY^d\bigr)\Bigr). \label{eq:3.2.3b}
\end{eqnarray}
These vectors are obtained from the position vectors $tX^a$ and $tX^a(w)$ 
in two steps: first the zero of the angle parameter $w$ was shifted, and 
then they were modified by the curvature terms. Clearly, the first is only 
a change of parameterization, and we could have parameterized the points 
$r_1$ and $r_2$ in this way even in Section \ref{sec-2}. The advantage of 
this re-parameterization is that $\xi^a_1$ and $\xi^a_2$ evidently have the 
same structure. However, the second step is essential here: it is this point 
where the effect of curvature is taken into account, viz. that the position 
vectors of the points $r_1$ and $r_2$, given in the flat geometry of the 
tangent space by $X^a(\pm w/2)$, are shifted by the curvature both with 
respect to $q$ and relative to each other. 

The norm of these vectors with respect to the \emph{flat} Minkowski metric 
$\eta_{ab}$ is 
\begin{eqnarray}
\eta_{ab}\xi^a_1\xi^b_1\!\!\!\!&=\!\!\!\!&\eta_{ab}\xi^a_2\xi^b_2=-t^2\Bigl(1-
  \sin^2\frac{w}{2}\frac{t^2}{3}R_{abcd}X^aY^bX^cY^d+O(w^2)O(t^4)\Bigr)=
  -t^2+\nonumber \\
\!\!\!\!&+\!\!\!\!&O(w^2)O(t^4)=-(D_{q,r_1})^2+O(w^2)O(t^4)=-(D_{q,r_2})^2+
  O(w^2)O(t^4), \label{eq:3.2.4a} \\ 
\eta_{ab}\bigl(\xi^a_2\!\!\!\!&-\!\!\!\!&\xi^a_1\bigr)\bigl(\xi^b_2-\xi^b_1
  \bigr)=-4\sin^2\frac{w}{2}\,t^2\Bigl(1-\frac{t^2}{3}R_{abcd}X^aY^bX^cY^d+
  O(t^4)\Bigr)= \nonumber \\
=\!\!\!\!&-\!\!\!\!&(D_{r_1,r_2})^2+O(t^2)O(w^3)+O(t^5)O(w^2).
  \label{eq:3.2.4c}
\end{eqnarray}
Hence, apart from higher order terms, the distances $D_{q,r_1}$, $D_{q,r_2}$ 
and $D_{r_1,r_2}$ given by (\ref{eq:3.2.1a})-(\ref{eq:3.2.1c}) can in fact be 
rewritten as the Minkowski norm of the vectors $\xi^a_1$, $\xi^a_2$ and 
$\xi^a_2-\xi^a_1$, respectively. Taking the derivatives of these norms we 
obtain 
\begin{eqnarray}
&{}&\frac{1}{2}\Bigl(\frac{\partial^2}{\partial t^2}\bigl(\eta_{ab}\xi^a_1
  \xi^b_1\bigr)\Bigr)_{t=w=0}=\frac{1}{2}\Bigl(\frac{\partial^2}{\partial t^2}
  \bigl(\eta_{ab}\xi^a_2\xi^b_2\bigr)\Bigr)_{t=w=0}=-1, \label{eq:3.2.5a}\\
&{}&\frac{1}{16}\Bigl(\frac{\partial^6}{\partial t^4\partial w^2}\bigl(\eta
  _{ab}(\xi^a_2-\xi^a_1)(\xi^b_2-\xi^b_1)\bigr)\Bigr)_{t=w=0}=R_{abcd}X^aY^bX^cY^d.
  \label{eq:3.2.5b}
\end{eqnarray}
Thus, in the $t,w\to0$ limit, the above norm of both $\xi^a_1$ and $\xi^a_2$ 
provides the \emph{physical distance} $t$ in the leading order along the 
radial geodesics $\alpha$; and the norm of $\xi^a_2-\xi^a_1$, also with 
respect to the \emph{flat} Minkowski metric $\eta_{ab}$, reproduces the 
component $R_{abcd}X^aY^bX^cY^d$ of the curvature tensor at $q$ (see Lemma 
\ref{lemma-2.2.1} and equation (\ref{eq:2.2.2})). The significance of this 
form of the distances will be clear in subsection \ref{sub-4.2}. 


\subsubsection{Distances between timelike geodesics}
\label{sub-3.2.2}

In the present subsection, we make the analogy with the Minkowski case even 
closer by reinterpreting the distances (\ref{eq:3.2.1a})-(\ref{eq:3.2.1c}) 
between the points $q$, $r_1$ and $r_2$ to be distances between certain 
\emph{timelike geodesics} through these points. 

If $r,r'\in U$ are any two spacelike separated points, and $\gamma$ and 
$\gamma'$ are arbitrary timelike geodesics through $r$ and $r'$, 
respectively, such that they are \emph{orthogonal} to the uniquely 
determined spacelike geodesic segment between $r$ and $r'$, then their 
distance $D(\gamma,\gamma')$ is \emph{defined} to be just $D_{r,r'}$. $\gamma$ 
and $\gamma'$ can always be chosen such that the tangent of $\gamma$, 
parallelly propagated from $r$ into $r'$ along the spacelike geodesic 
segment, is not parallel with that of $\gamma'$. (We emphasize that, in 
contrast to the Minkowski case, the distance $D$ is \emph{not} defined 
between \emph{any} two timelike geodesics, not even in a convex normal 
neighbourhood.) If $r=r'$, i.e. when these geodesics intersect each other, 
the distance between them is clearly zero. On the other hand, even if $r$ 
and $r'$ are different and spacelike separated, and hence the distance 
between the geodesics $\gamma$ and $\gamma'$ defined in this way is non-zero, 
then these geodesics may still intersect each other somewhere in the common 
part of the chronological future or past of the two points, $I^+(r)\cap I^+
(r')$ or $I^-(r)\cap I^-(r')$ (for the notation see e.g. \cite{Pe72,HE}). 
This phenomenon is due to geodesic focusing. However, this intersection 
should be \emph{outside} $U$. Hence, the intersection of two timelike 
geodesics, i.e. the points of the neighbourhood $U$, can be characterized 
\emph{locally} by the vanishing of their distance, yielding locally an 
\emph{operational definition} of spacetime points using only the notion of 
their distance\footnote{
Without restricting our considerations to convex normal neighbourhoods, to 
have an analogous characterization of the spacetime points by the set of 
the future/past directed timelike curves through them the future/past 
distinguishing condition, a rather weak global causality condition, must 
also be satisfied \cite{Pe72,HE}.}. 
If two timelike geodesics do not intersect each other in $U$, then either 
their distance is not defined, or, if it is defined, then it is not zero. 
Thus the restriction of the notion of $D$ to certain pairs of timelike 
geodesics in a small enough $U$ made it possible to characterize the 
points of $U$ by $D$. 

The above considerations, together with the construction made at the end 
of the previous subsection, motivate the following geometric setup. 
Let $p^\alpha_{01}$ and $p^\alpha_{02}$ be different future pointing timelike 
vectors at $q$ which are orthogonal to $X^\alpha$ and $Y^\alpha$. By parallel 
transport along $\alpha_0$ these vectors can be propagated from $q$ into 
$r_1$, and by parallel transport along $\alpha_w$ into $r_2$. Then let us 
choose two different future pointing timelike vectors $p^\alpha_{11}$ and 
$p^\alpha_{12}$ at $r_1$ which are orthogonal to the spacelike geodesics 
$\alpha_0$ and $\chi$ there. Moreover, since $4=\dim M>3$, $p^\alpha_{11}$ 
and $p^\alpha_{12}$ can always be chosen not to be parallel with any of 
$p^\alpha_{01}$ and $p^\alpha_{02}$ propagated here. By parallel transport, we 
propagate $p^\alpha_{11}$ and $p^\alpha_{12}$ along $\chi$ from $r_1$ into $r_2$. 
In a similar way, we choose $p^\alpha_{21}$ and $p^\alpha_{22}$ to be different 
future pointing timelike vectors at $r_2$ which are orthogonal to $\alpha_w$ 
and $\chi$, and which are not parallel with any of $p^\alpha_{01}$, $p^\alpha
_{01}$, $p^\alpha_{21}$ and $p^\alpha_{22}$ propagated there. Again, since $\dim M
=4$, $p^\alpha_{21}$ and $p^\alpha_{22}$ can be chosen in this manner. 

Then there are future directed timelike geodesics $\gamma_{01}$ and 
$\gamma_{02}$ through $q$ with tangents $p^\alpha_{01}$ and $p^\alpha_{02}$, 
timelike geodesics $\gamma_{11}$ and $\gamma_{12}$ through $r_1$ with 
tangents $p^\alpha_{11}$ and $p^\alpha_{12}$, and timelike geodesics $\gamma
_{21}$ and $\gamma_{22}$ through $r_2$ with tangents $p^\alpha_{21}$ and 
$p^\alpha_{22}$, respectively. The distance between them are \emph{defined} 
simply to be 
\begin{eqnarray}
&{}&D(\gamma_{01},\gamma_{02})=D(\gamma_{11},\gamma_{12})=D(\gamma_{21},
  \gamma_{22})=0, \label{eq:3.2.6a} \\
&{}&D(\gamma_{01},\gamma_{11})=D(\gamma_{01},\gamma_{12})=D(\gamma_{02},
  \gamma_{11})=D(\gamma_{02},\gamma_{12})=D_{q,r_1}, \label{eq:3.2.6b} \\
&{}&D(\gamma_{01},\gamma_{21})=D(\gamma_{01},\gamma_{22})=D(\gamma_{02},
  \gamma_{21})=D(\gamma_{02},\gamma_{22})=D_{q,r_2}, \label{eq:3.2.6c} \\
&{}&D(\gamma_{11},\gamma_{21})=D(\gamma_{11},\gamma_{22})=D(\gamma_{12},
  \gamma_{21})=D(\gamma_{12},\gamma_{22})=D_{r_1,r_2}. \label{eq:3.2.6d}
\end{eqnarray}
With these definitions we reinterpret the above geometric setup in the 
following way: 

Let us consider the above six timelike geodesics as world lines of massive 
classical point particles with the given 4-momenta as tangents and let us 
consider the distances (\ref{eq:3.2.6a})-(\ref{eq:3.2.6d}) between them to 
be \emph{primarily given}. Then we may say that, by (\ref{eq:3.2.6a}), they 
\emph{define} three points, and let us call them $q$, $r_1$ and $r_2$. The 
distance between these points is \emph{defined} by the primarily given 
distances (\ref{eq:3.2.6b})-(\ref{eq:3.2.6d}) between the world lines. Then 
equation (\ref{eq:2.2.1}) in Lemma \ref{lemma-2.2.1} tells us how the 
component $R_{abcd}X^aY^bX^cY^d$ of the curvature tensor at the point $q$ can 
be recovered, or, alternatively, be \emph{defined}, by the distances between 
these timelike geodesics. Then by (\ref{eq:2.1.1a}) and (\ref{eq:2.1.1b}) 
the $C^2$ metric tensor can be recovered on $U$. The genericness of the 
curvature at $q$ ensures the genericness of the metric on $U$. It is this 
classical picture that will be behind the quantum mechanical setup that we 
use to get our main results in Section \ref{sec-4}. 


\section{Lorentzian 4-geometries from quantum mechanical empirical distances}
\label{sec-4}

\subsection{The empirical distance}
\label{sub-4.1}

The composite quantum mechanical system considered in Theorem \ref{th-3.1.1} 
is the formal union of any large $N$ number of independent $E(1,3)$-invariant 
elementary quantum mechanical systems. (Recall that a quantum system in its
algebraic formulation is thought to be specified completely if the algebra of 
its observables and the representation of this algebra on a complex separable 
Hilbert space are fixed. Such a system is called an $E(1,3)$-invariant 
\emph{elementary} quantum mechanical system if the algebra is the universal 
enveloping algebra of the Lie algebra $e(1,3)$ of $E(1,3)$ and the 
representation is a unitary irreducible representation. See e.g. 
\cite{NeWign,StWi}.) The Hilbert space of its vector states is the tensor 
product ${\cal H}:={\cal H}_1\otimes\cdots\otimes{\cal H}_N$ of the Hilbert 
spaces of the elementary subsystems, and the Hilbert spaces ${\cal H}_{\bi}$, 
${\bi}=1,...,N$, are carrier spaces of unitary irreducible representations 
of $E(1,3)$. The states in Theorem \ref{th-3.1.1} are the tensor products of 
the states of the elementary systems with the form $\phi_{\bi}$, or in the 
bra-ket notation $\vert\phi_{\bi}\rangle$, given by 
\begin{equation}
\vert\phi_{\bi}\rangle=\exp\bigl(\frac{\rm i}{\hbar}p_{{\bi}e}\xi^e_{\bi}\bigr)
{\bf U}_{\bi}\vert\psi_{\bi}\rangle. \label{eq:4.1.1}
\end{equation}
Here $\vert\psi_{\bi}\rangle=\vert\psi_{s_{\bi},s_{\bi}}\rangle$ are special, 
co-moving centre-of-mass states in the unitary, irreducible representation 
of the quantum mechanical Poincar\'e group $E(1,3)$ with positive rest mass 
$\mu_{\bi}$ and spin $s_{\bi}$, $p^e_{\bi}$ is the 4-momentum (as multiplication 
operators) in this representation, ${\bf U}_{\bi}$ is the unitary operator 
that implements the Lorentz boost $\Lambda^a_{\bi}{}_b$ appearing in the 
classical expression $\gamma^a_{\bi}(u)=\Lambda^a_{\bi}{}_b\gamma^b_0(u)+\xi^a
_{\bi}$ of the timelike straight line $\gamma^a_{\bi}$, and $\xi^a_{\bi}$ is 
this translation (for the details see \cite{Sz23a}, and, for the 
representation, its Appendix A.2). The states $\vert\psi_{\bi}\rangle$ play 
the role analogous to that of the (reference) world line $\gamma^a(u)=u\,
\delta^a_0$ given in the Cartesian coordinates in Minkowski space. 
(Recall that a state $\vert\psi\rangle$ is called a special, co-moving 
centre-of-mass state if, in this state, the expectation value of the 
centre-of-mass vector operator is vanishing, the expectation value of the 
energy-momentum 4-vector has only the time component and the expectation 
value of the Pauli--Lubanski spin has only a single spacelike component. 
Such states were constructed \emph{explicitly} in \cite{Sz23a}.) In the 
classical limit, which is defined by a sequence of states $\phi_{{\bi}k}$, 
$k\in\mathbb{N}$ (and denoted by $\Clim$), the two Casimir invariants 
$\mu_{\bi}$ and $s_{\bi}$ are linked via $\mu_{\bi}=O(s_{\bi})$ and the limit 
itself is defined by $s_{\bi}\to\infty$. Thus, it is essentially the smallest 
of the spins $s_1,...,s_N$ that plays the role of the index $k$ in Theorem 
\ref{th-3.1.1}. (For the details, see \cite{Sz23a}). 

The (square of the) empirical distance $d_{\bi\bj}$ between the ${\bi}$th and 
${\bj}$th elementary subsystems of the composite system in the state 
$\phi:=\phi_1\otimes\cdots\otimes\phi_N$ was defined by 
\begin{equation}
d^2_{\bi\bj}:=\frac{\langle\phi\vert(\Sigma^a_{\bi\bj}\varepsilon_{acde}p^c_{\bi}
p^d_{\bj})(\varepsilon^e{}_{ghb}p^g_{\bi}p^h_{\bj}\Sigma^b_{\bi\bj})\vert\phi
\rangle}{\langle\phi\vert\bigl({\bf P}^4_{\bi\bj}-\mu^2_{\bi}\mu^2_{\bj}{\bf I}_1
\otimes\cdots\otimes{\bf I}_N\bigr)\vert\phi\rangle}, 
\label{eq:4.1.2}
\end{equation}
where ${\bf I}_{\bi}$ is the identity operator on ${\cal H}_{\bi}$, and, for 
${\bi}<{\bj}$, 
\begin{eqnarray}
&{}&\Sigma^a_{\bi\bj}:=\frac{1}{\mu^2_{\bi}}{\bf I}_1\otimes\cdots\otimes
  {\bf C}^a_{\bi}\otimes\cdots\otimes{\bf I}_N-\frac{1}{\mu^2_{\bj}}{\bf I}_1
  \otimes\cdots\otimes{\bf C}^a_{\bj}\otimes\cdots\otimes{\bf I}_N, 
  \label{eq:4.1.3a} \\
&{}&{\bf P}^2_{\bi\bj}:=\eta_{ab}{\bf I}_1\otimes\cdots\otimes{\bf p}^a_{\bi}
  \otimes\cdots\otimes{\bf p}^b_{\bj}\otimes\cdots\otimes{\bf I}_N.
  \label{eq:4.1.3b}
\end{eqnarray}
Here ${\bf C}^a_{\bi}$, the (self-adjoint) centre-of-mass operators, are built 
from the 4-momentum and angular momentum tensor operators of the elementary 
systems according to the general definition ${\bf C}^a:={\bf J}^{ab}{\bf p}_b
-\frac{3}{2}{\rm i}\hbar{\bf p}^a$. The structure of (\ref{eq:4.1.2}) was 
motivated by the structure of the classical expression (\ref{eq:3.1.3}). The 
role of the denominator in (\ref{eq:4.1.2}) is simply the normalization of 
the projection $(\varepsilon^a{}_{cde}p^c_{\bi}p^d_{\bj})(\varepsilon^e{}_{ghb}
p^g_{\bi}p^h_{\bj})$. (See the classical expression (\ref{eq:3.1.2}), and also 
\cite{Sz21c} and \cite{Sz22a} in the $SU(2)$ and $E(3)$-invariant cases, 
respectively, where the `empirical' quantum physical quantities considered 
there have the similar structure). 

The structure (\ref{eq:4.1.3a}) of $\Sigma^a_{\bi\bj}$ and (\ref{eq:4.1.3b}) 
of ${\bf P}^2_{\bi\bj}$ implies that, by the \emph{pure tensor product} nature 
of the state $\phi$, $d^2_{\bi\bj}$ depends \emph{only} on the states $\phi
_{\bi}$ and $\phi_{\bj}$ of the respective elementary systems, and it is 
independent of the state of the other subsystems. In particular, in the 
states of the form (\ref{eq:4.1.1}), $d^2_{\bi\bj}$ gives $-\Pi_{ab}(\xi^a_{\bi}
-\xi^a_{\bj})(\xi^b_{\bi}-\xi^b_{\bj})$ in the classical limit with 
asymptotically vanishing uncertainty\footnote{
Since by (\ref{eq:3.1.2}) $d^2_{\bi\bj}$ is the quotient of two expectation 
values rather than the expectation value of a single self-adjoint operator, 
its uncertainty is \emph{not} the standard deviation of an operator in the 
given state. Essentially, this uncertainty is the sum of the standard 
deviation of the operator in the numerator and of the operator in the 
denominator in (\ref{eq:3.1.2}). For the details, see \cite{Sz23a}.}, 
moreover, if the vectors $\xi^a_{\bi}$ and $\xi^a_{\bi}$ are chosen to be the 
translations in the parameterization of the straight lines $\gamma^a_{\bi}$ 
and $\gamma^a_{\bj}$ in Minkowski space, then this limit is just the square 
of the correct Lorentzian distance $D_{\bi\bj}$ between $\gamma^a_{\bi}$ and 
$\gamma^a_{\bj}$: $\Clim d^2_{\bi\bj}=D^2_{\bi\bj}$ (see (\ref{eq:3.1.1}) and 
Theorem \ref{th-3.1.1} above). This is the way how the metric structure of 
Minkowski space could be recovered from $d^2_{\bi\bj}$. 

As we already noted in (the first footnote in subsection 1.2 of) 
\cite{Sz23a}, the root of the boost-rotation and the translation symmetries 
in the Minkowski \emph{affine} space, introduced as the dual to the momentum 
space, is different. The boost-rotation symmetries are rooted in the 
intrinsic Minkowski \emph{vector space} structure of the momentum space, 
while the translations come from the \emph{assumption} that changing the 
wave function not only by a constant, but even by a special 
momentum-dependent phase factor, $\exp({\rm i}p_e\xi^e/\hbar)$, as a symmetry 
of the quantum mechanical system, came from a metric symmetry of the 
Minkowski space, too. It is this assumption that makes the Minkowski 
\emph{vector space} to be Minkowski \emph{affine space}. Moreover, in 
subsection \ref{sub-2.1} we saw that the Lorentzian symmetry of the Minkowski 
vector space survives in curved spacetimes as an exact symmetry in the 
tangent spaces, and hence as a symmetry that takes Riemannian normal 
coordinate systems into such coordinate systems. On the other hand, due to 
the curvature, the naive translation of the Riemannian normal coordinates, 
$x^a\mapsto x^a+\xi^a$, does \emph{not} yield Riemannian normal coordinates. 
Thus the Lorentz and translation symmetries of the Minkowski affine space 
separate even further in the presence of curvature. Therefore, it seems 
natural to search for the appropriate states in (\ref{eq:4.1.2}) among 
those in which only the translations, or rather the position vectors, 
deviate from the states yielding the flat spacetime distances. 

In the next subsection we show that the local metric structure of 
\emph{curved} spacetimes can also be recovered from, or rather 
\emph{defined} by, \emph{tensor product states} simply by modifying the 
position vectors in (\ref{eq:3.1.1}) appropriately. 


\subsection{Recovering the Riemann tensor from pure tensor product states}
\label{sub-4.2}

We can apply the general ideas above to the composite system consisting of 
six elementary subsystems indexed by ${\bi}=01,02,11,12,21,22$ and whose 
state is the tensor product state 
\begin{eqnarray}
\vert\phi\rangle:=\!\!\!\!\!&{}\!\!\!\!\!&\exp\Bigl(\frac{\rm i}{\hbar}
  \bigl(p_{01e}+p_{02e}\bigr)\xi^e_{01}\Bigr)\Bigl({\bf U}_{01}\vert\psi_{01}
  \rangle\otimes{\bf U}_{02}\vert\psi_{02}\rangle\Bigr) \nonumber \\
\otimes\!\!\!\!\!&{}\!\!\!\!\!&\exp\Bigl(\frac{\rm i}{\hbar}\bigl(p_{11e}
  +p_{12e}\bigr)\xi^e_{11}\Bigr)\Bigl({\bf U}_{11}\vert\psi_{11}\rangle
  \otimes{\bf U}_{12}\vert\psi_{12}\rangle\Bigr) \nonumber \\
\otimes\!\!\!\!\!&{}\!\!\!\!\!&\exp\Bigl(\frac{\rm i}{\hbar}\bigl(p_{21e}
  +p_{22e}\bigr)\xi^e_{21}\Bigr)\Bigl({\bf U}_{21}\vert\psi_{21}\rangle\otimes
  {\bf U}_{22}\vert\psi_{22}\rangle\Bigr). 
\label{eq:4.2.1}
\end{eqnarray}
Note that the position vectors in the subsystems $01$ and $02$, in $11$ and 
$12$, and in $21$ and $22$ are the same. If the Lorentz transformations that 
are represented by the unitary operators in (\ref{eq:4.2.1}) are all 
different, then by the discussion in subsection \ref{sub-3.2.1} the 
classical limit, $\Clim d^2_{\bi\bj}$, is well defined for any $\bi,\bj$. 
Moreover, $\Clim d_{01,02}=\Clim d_{11,12}=\Clim d_{21,22}=0$ (see 
(\ref{eq:3.1.1})). Thus, the role of the pairs $(01,02)$, $(11,12)$ and 
$(21,22)$ of the subsystems with the given structure of their states is to 
\emph{define} three points in the classical limit, and we call these points 
$q$, $r_1$ and $r_2$, respectively. 

However, in general the classical limit of the empirical distances 
$d_{01,11}$, $d_{01,12}$, $d_{02,11}$ and $d_{02,12}$ are all different unless 
the Lorentz boosts $\Lambda^a_{01}{}_0$, $\Lambda^a_{02}{}_0$, $\Lambda^a_{11}
{}_0$ and $\Lambda^a_{12}{}_0$ are all orthogonal to $\xi^a_{11}-\xi^a_{01}$ 
(see equation (\ref{eq:3.1.5})). In fact, although the classical limit of 
the square of these empirical distances has the form $-\Pi_{ab}(\xi^a_{11}-
\xi^a_{01})(\xi^b_{11}-\xi^b_{01})$ (see equation (\ref{eq:3.1.1})), but the 
projection $\Pi_{ab}$ in these four cases are \emph{different} (see 
equation (\ref{eq:3.1.2})): it is the projection to the spacelike 2-space 
orthogonal to $\Lambda^a_{01}{}_0$ and $\Lambda^a_{11}{}_0$, to $\Lambda^a
_{01}{}_0$ and $\Lambda^a_{12}{}_0$, to $\Lambda^a_{02}{}_0$ and $\Lambda^a
_{11}{}_0$, and to $\Lambda^a_{02}{}_0$ and $\Lambda^a_{12}{}_0$, respectively. 
These spacelike 2-spaces coincide, and hence $\Clim d_{01,11}=\Clim d_{01,12}
=\Clim d_{02,11}=\Clim d_{02,12}$ holds, if the orthogonality conditions in 
(\ref{eq:3.1.5}) are satisfied. In this case the common classical limit of 
their square is 
\begin{equation}
\Clim d^2_{01,11}=\Clim d^2_{01,12}=\Clim d^2_{02,11}=\Clim d^2_{02,12}=-\eta_{ab}
(\xi^a_{11}-\xi^a_{01})(\xi^b_{11}-\xi^b_{01}) \label{eq:4.2.2}
\end{equation}
(see the structure (\ref{eq:3.2.2}) of the projection $\Pi_{ab}$), which is 
just the square of the Minkowski norm of $\xi^a_{11}-\xi^a_{01}$. We have 
analogous conditions on the equality of the classical limit of $d_{01,21}$, 
$d_{01,22}$, $d_{02,21}$ and $d_{02,22}$ in terms of $\xi^a_{21}-\xi^a_{01}$ and 
the Lorentz boosts $\Lambda^a_{01}{}_0$, $\Lambda^a_{02}{}_0$, $\Lambda^a_{21}
{}_0$ and $\Lambda^a_{22}{}_0$; and also on the equality of the classical 
limit of $d_{11,21}$, $d_{11,22}$, $d_{12,21}$ and $d_{12,22}$ in terms of $\xi^a
_{21}-\xi^a_{11}$ and $\Lambda^a_{11}{}_0$, $\Lambda^a_{12}{}_0$, $\Lambda^a_{21}
{}_0$ and $\Lambda^a_{22}{}_0$. Therefore, if (\ref{eq:3.1.5}) and the 
analogous two conditions are satisfied, i.e. all the six boosts $\Lambda^a
_{01}{}_0$, $\Lambda^a_{02}{}_0$, $\Lambda^a_{11}{}_0$, $\Lambda^a_{12}{}_0$, 
$\Lambda^a_{21}{}_0$, $\Lambda^a_{22}{}_0$ are orthogonal to the spacelike 
2-plane spanned by $\xi^a_{11}-\xi^a_{01}$ and $\xi^a_{21}-\xi^a_{01}$, then 
the resulting three distances in the classical limit can be interpreted as 
the distances between the points $q$, $r_1$ and $r_2$ above like in 
(\ref{eq:3.2.6b})-(\ref{eq:3.2.6d}), and the distances themselves are given 
by the norm of $\xi^a_{11}-\xi^a_{01}$, $\xi^a_{21}-\xi^a_{01}$ and $\xi^a_{21}-
\xi^a_{11}$ with respect to the \emph{flat} metric $\eta_{ab}$. Note that, so 
far, the position vectors $\xi^a_{01}$, $\xi^a_{11}$ and $\xi^a_{21}$ have not 
been specified. 

If the position vectors $\xi^a_{01}$, $\xi^a_{11}$ and $\xi^a_{21}$ were chosen 
to be those in the parametrization of the timelike straight lines in 
Minkowski spacetime, then these distances would be the \emph{flat} spacetime 
distances. In the rest of the present subsection, we show that, choosing 
these position vectors according to the results of subsection \ref{sub-3.2.1}, 
we can recover the local metric structure of any \emph{curved} Lorentzian 
spacetime as well. 

Choosing the vectors $\xi^a_{11}$ and $\xi^a_{12}$ in (\ref{eq:4.2.1}) to be 
\begin{equation}
\xi^a_{11}:=\xi^a_{01}+\xi^a_1, \hskip 20pt \xi^a_{21}:=\xi^a_{01}+\xi^a_2,
\label{eq:4.2.3}
\end{equation}
where the vectors $\xi^a_1$ and $\xi^a_2$ are the \emph{shifted relative 
position vectors} given by (\ref{eq:3.2.3a}) and (\ref{eq:3.2.3b}), 
respectively, we obtain a two-parameter family of tensor product states, viz. 
\begin{eqnarray}
\vert\phi(t,w)\rangle=\exp\Bigl(\!\!\!\!\!&{}\!\!\!\!\!&\frac{\rm i}{\hbar}
(p_{01e}+p_{02e})\xi^e_{01}+ \nonumber \\
+\!\!\!\!\!&{}\!\!\!\!\!&\frac{\rm i}{\hbar}(p_{11e}+p_{12e})\bigl(\xi^e_{01}
  +tX^e(-\frac{w}{2})+\sin\frac{w}{2}\frac{t^3}{6}R^e{}_{bcd}X^bX^cY^d\bigr)+
  \nonumber \\
+\!\!\!\!\!&{}\!\!\!\!\!&\frac{\rm i}{\hbar}(p_{21e}+p_{22e})\bigl(\xi^e_{01}+
  tX^e(\frac{w}{2})-\sin\frac{w}{2}\frac{t^3}{6}R^e{}_{cdb}X^cX^dY^b\bigr)
  \Bigr)\vert\chi\rangle, \label{eq:4.2.4a}
\end{eqnarray}
where 
\begin{equation}
\vert\chi\rangle:={\bf U}_{01}\vert\psi_{01}\rangle\otimes{\bf U}_{02}\vert
\psi_{02}\rangle\otimes{\bf U}_{11}\vert\psi_{11}\rangle\otimes{\bf U}_{12}
\vert\psi_{12}\rangle\otimes{\bf U}_{21}\vert\psi_{21}\rangle\otimes
{\bf U}_{22}\vert\psi_{22}\rangle. \label{eq:4.2.4b}
\end{equation}
As in (\ref{eq:4.2.1}), here the states $\vert\psi_{\bi}\rangle$, ${\bi}=01,
02,11,12,21,22$, are the special co-moving centre-of-mass states in the 
unitary irreducible representation of $E(1,3)$ labeled by the Casimir 
invariants $\mu_{\bi}$ and $s_{\bi}$. The Lorentz boosts, represented by the 
unitary operators in (\ref{eq:4.2.4b}), are chosen to satisfy the 
orthogonality conditions discussed at the end of the second paragraph above. 
If the states $\vert\psi_{\bi}\rangle$ are chosen from a sequence $\vert\psi
_{{\bi}k}\rangle$, $k\in\mathbb{N}$, of special co-moving centre-of-mass 
states that define the classical limit, then $d^2_{\bi\bj}$ can be evaluated 
in these states and one can take the classical limit $k:={\rm min}\{s_{01},
...,s_{22}\}\to\infty$. Then, repeating the argumentation behind Theorem 
\ref{th-3.1.1} and using (\ref{eq:3.2.4a})-(\ref{eq:3.2.4c}), we obtain 
\begin{eqnarray}
\Clim d^2_{01,02}=\Clim d^2_{11,12}=\Clim d^2_{21,22}\!\!\!\!&=\!\!\!\!&0,
  \label{eq:3.2.8a} \\
\Clim d^2_{01,11}=\Clim d^2_{01,12}=\Clim d^2_{02,11}=\Clim d^2_{02,12}\!\!\!\!&
  =\!\!\!\!&(D_{q,r_1})^2+O(t^4)O(w^2), \label{eq:3.2.8b} \\
\Clim d^2_{01,21}=\Clim d^2_{01,22}=\Clim d^2_{02,21}=\Clim d^2_{02,22}\!\!\!\!&
  =\!\!\!\!&(D_{q,r_2})^2+O(t^4)O(w^2), \label{eq:3.2.8c} \\
\Clim d^2_{11,21}=\Clim d^2_{11,22}=\Clim d^2_{12,21}=\Clim d^2_{12,22}\!\!\!\!&
  =\!\!\!\!&(D_{r_1,r_2})^2+ \nonumber \\
\!\!\!\!&+\!\!\!\!&O(t^2)O(w^3)+O(t^5)O(w^2). \label{eq:3.2.8d}
\end{eqnarray}
Hence, up to higher order corrections, the classical distances 
(\ref{eq:3.2.6a})-(\ref{eq:3.2.6d}) have been recovered from the quantum 
mechanical empirical distance $d^2_{\bi\bj}$ in the classical limit. With this 
result and as a consequence of Theorem \ref{th-3.1.1}, we have proven the 
next statement. 

\begin{theorem}\label{th-4.2.1}
Let $(M,g_{\alpha\beta})$ be a spacetime manifold, $q\in M$, and let $X^\alpha,
Y^\alpha\in T_qM$ be any two spacelike unit vectors that are orthogonal to 
each other. Let $r_1:=\exp_q(tX)$ and $r_2:=\exp_q(tX(w))$, where $X^\alpha
(w):=\cos wX^\alpha+\sin wY^\alpha$, and let $D_{q,r_1}$, $D_{q,r_2}$ and 
$D_{r_1,r_2}$ be the distances between these points.
Then there exist six Poincar\'e-invariant quantum mechanical systems 
${\cal S}_{\bi}$, ${\bi}=01,02,11,12,21,22$, and a sequence $\phi_{{\bi}k}
(t,w)$, $k\in\mathbb{N}$, of two-parameter families of their vector states, 
$0\leq t,w<\epsilon$ for some $\epsilon>0$, such that in the $k\to\infty$ 
limit the square of the empirical distances, $d^2_{\bi\bj}$, calculated in the 
tensor product states $\phi_{01k}(t,w)\otimes\cdots\otimes\phi_{22k}(t,w)$, 
tend with asymptotically vanishing uncertainty to the values given by 
(\ref{eq:3.2.8a})-(\ref{eq:3.2.8d}). 
\end{theorem}
Thus, as a consequence of (\ref{eq:3.2.5a})-(\ref{eq:3.2.5b}), the 
component $R_{\alpha\beta\gamma\delta}X^\alpha Y^\beta X^\gamma Y^\delta$ of the 
curvature tensor at $q$ has been recovered. In the Appendix, we show that 
there are 20 \emph{spacelike} 2-planes in $T_qM$ such that the 
corresponding sectional curvatures determine all the independent components 
of the curvature tensor. Therefore, choosing the vectors $X^\alpha,Y^\alpha
\in T_qM$ to span these twenty spacelike 2-planes, all the components of 
the curvature tensor can be determined, or rather \emph{defined}, in the 
classical limit from the empirical distances between the constituents of 
sextets of Poincar\'e-invariant quantum mechanical systems via Theorem 
\ref{th-4.2.1}. Since by equations (\ref{eq:2.1.1a})-(\ref{eq:2.1.1b}) the 
whole $C^2$ metric on $U$ is determined by the curvature and the curvature is 
arbitrary, \emph{the local geometry of any curved $C^2$ Lorentzian spacetime 
can be determined, or rather defined, in the classical limit by observables 
of abstract, Poincar\'e-invariant quantum mechanical systems}.

\section{Final remarks}
\label{sec-5}

The strategies to resolve the well known conflicts between general relativity 
and quantum theory are usually based on one of the two paradigms: first, 
gravity is `fundamental' (like electromagnetism), and hence it must be the 
subject of some (more or less standard) active quantization procedure; and 
the other is that gravity is `emergent' (like thermo- or hydrodynamics), in 
which case gravity should be derivable from (probably some modified, or 
rather `improved') quantum theory. In both cases one must have some extra, 
still not justified \emph{a priori} assumption on the existing theories. 
Nevertheless, according to Wheeler's `radical conservatism' \cite{Th}, 
before introducing a new paradigm to explain some phenomenon or to resolve 
some difficulty, we should exhaust all the possibilities that our existing, 
known theories provide. It could be the ultimate failure of all these 
attempts that could indicate what kind of new paradigm (if any) should be 
adopted. 

Our present investigations (and, in fact, the previous ones in 
\cite{Sz21c,Sz22a,Sz23a}, too) were done in this spirit. We took general 
relativity and quantum mechanics as they are, and we found that the metric 
structure of the Euclidean 3-space and of the flat and curved spacetimes 
can be derived from quantum mechanics \emph{strictly in the framework 
provided by these two theories}. A by-product of these investigations is 
that, in contrast to general expectations, these metric structures could be 
derived from \emph{pure tensor product states} of the quantum mechanical 
subsystems, i.e. without any entanglement of them. It is the observables 
that are entangled. 

While the preparation of the present paper was in its final stage, the paper 
\cite{Matsuda} appeared with the idea similar to that we are proposing here, 
viz. that the quantum theory should live in the flat geometry of the tangent 
spaces of curved spacetimes, and the quantum physical phenomena on a 
neighbourhood of a spacetime point, rather than only on the whole spacetime, 
should also be investigated in this way. However, the mathematical/technical 
realization of this idea appears to be rather different in \cite{Matsuda} 
and in the present paper. 

No funds, grants or support was received. 


\appendix 

\section{Appendix: Curvature from purely spatial 2-surfaces}
\label{sec-A}

In this appendix, we show that although the spacetime metric is Lorentzian, 
all the components of the curvature tensor can be recovered from the 
sectional curvatures determined by \emph{purely spacelike} 2-surfaces. 

If $X^\alpha,Y^\alpha\in T_qM$ are any two linearly independent vectors, then 
let $[X.Y]$ denote the 2-plane in $T_qM$ spanned by $X^\alpha$ and $Y^\alpha$ 
(which should not be confused with the Lie bracket of them), and form 
$A(X,Y):=(g_{\alpha\beta}X^\alpha X^\beta)(g_{\gamma\delta}Y^\gamma Y^\delta)-
(g_{\alpha\beta}X^\alpha Y^\beta)^2$. If the 2-plane $[X,Y]$ is non-degenerate 
(in the sense that $A(X,Y)\not=0$), then the sectional curvature 
corresponding to the 2-plane $[X,Y]$ is defined by 
\begin{equation}
K(X,Y):=\frac{R_{\alpha\beta\gamma\delta}X^\alpha Y^\beta X^\gamma Y^\delta}{A(X,Y)}.
\label{eq:A.1.1}
\end{equation}
(In Lorentzian geometries, $[X.Y]$ is degenerate if e.g. $X^\alpha$ is null 
and $Y^\alpha$ is a spacelike vector orthogonal to $X^\alpha$.) Clearly, 
$K(X,Y)=K(Y,X)$, and it depends only on the (non-degenerate) 2-plane 
$[X,Y]$, but it is independent of the actual vectors $X^\alpha$ and $Y^\alpha$ 
that span $[X,Y]$. In particular, $K(X,aX+bY)=K(X,Y)$ holds for any real $a$ 
and non-zero real $b$. The significance of the sectional curvatures is that, 
as it is proven e.g. in \cite{Milnor,Spivak2}, they determine the curvature 
tensor at $q$ completely. 

However, we need more than the general proof above: we want the 
\emph{explicit expression} of the components of the curvature tensor in 
terms of the sectional curvatures. As one can check e.g. by direct 
calculations, this is given by 
\begin{eqnarray}
R_{\alpha\beta\gamma\delta}\!\!\!\!\!&{}\!\!\!\!\!&X^\alpha Y^\beta Z^\gamma W^\delta
  = \label{eq:A.1.2} \\
=\frac{1}{6}\Bigl\{\!\!\!\!\!&{}\!\!\!\!\!&A(X+Z,Y+W)K(X+Z,Y+W)
  -A(Y+Z,X+W)K(Y+Z,X+W)+ \nonumber \\
+\!\!\!\!\!&{}\!\!\!\!\!&A(Y+Z,W)K(Y+Z,W)+A(Y+Z,X)K(Y+Z,X)- \nonumber \\
-\!\!\!\!\!&{}\!\!\!\!\!&A(X+Z,W)K(X+Z,W)-A(X+Z,Y)K(X+Z,Y)+ \nonumber \\
+\!\!\!\!\!&{}\!\!\!\!\!&A(X+W,Z)K(X+W,Z)+A(X+W,Y)K(X+W,Y)- \nonumber \\
-\!\!\!\!\!&{}\!\!\!\!\!&A(Y+W,Z)K(Y+W,Z)-A(Y+W,X)K(Y+W,X)+ \nonumber \\
+\!\!\!\!\!&{}\!\!\!\!\!&A(Y,Z)K(Y,Z)+
  A(W,X)K(W,X)-A(X,Z)K(X,Z)-A(W,Y)K(W,Y)\Bigr\}. \nonumber
\end{eqnarray}
This formula is valid even in any pseudo-Riemannian geometries, but here 
all the 2-planes must be non-degenerate. 

Since our aim is to link the independent components of the curvature tensor 
to \emph{purely spatial distances} via Lemma \ref{lemma-2.2.1}, in 
(\ref{eq:A.1.2}) we should use only \emph{spacelike} 2-planes and the 
corresponding sectional curvatures. However, if we used the vectors of the 
orthonormal basis $\{E^\alpha_a\}$ as the vectors $X^\alpha$, $Y^\alpha$, 
$Z^\alpha$ and $W^\alpha$ in (\ref{eq:A.1.2}), then some of the 2-planes would 
necessarily be timelike. Thus we choose a new basis in $T_qM$ consisting of 
purely spacelike vectors. 

Let us introduce the basis $\{E^\alpha_{\ua}\}=\{E^\alpha_1,E^\alpha_2,E^\alpha_3,
E^\alpha_4\}$, ${\ua}=1,...,4$, where  $E^\alpha_4:=\alpha E^\alpha_0+\beta E
^\alpha_1+\gamma E^\alpha_2+\delta E^\alpha_3$ and $\alpha,\beta,\gamma,\delta
\in\mathbb{R}$ for which $\alpha>0$ and $1+\alpha^2=\beta^2+\gamma^2+
\delta^2$ hold. Thus, the \emph{underlined} small Latin indices are 
referring to this basis. Clearly, $E^\alpha_4$ is a spacelike unit vector, 
but this is not orthogonal to the other vectors of the basis. We show that 
for appropriately chosen coefficients $\beta$, $\gamma$, $\delta$ all the 
2-planes that appear in the expression (\ref{eq:A.1.2}) of the components 
of the curvature tensor in the basis $\{E^\alpha_{\ua}\}$ are spacelike. 

If $W^\alpha=Y^\alpha$, then equation (\ref{eq:A.1.2}) reduces to 
\begin{equation}
R_{\alpha\beta\gamma\delta}X^\alpha Y^\beta Z^\gamma Y^\delta=\frac{1}{2}\Bigl(
A(X+Z,Y)K(X+Z,Y)-A(Z,Y)K(Z,Y)-A(X,Y)K(X,Y)\Bigr). \label{eq:A.1.3}
\end{equation}
18 of the 20 algebraically independent components of the curvature tensor 
are given by (\ref{eq:A.1.3}). Explicitly, since the vectors $E^\alpha_i$, 
$i=1,2,3$, form an orthonormal basis in a spacelike 3 dimensional subspace 
in $T_qM$, any two of these vectors span a spacelike 2-plane $[E_i,E_j]$, 
where $i\not=j$. Hence, for $i\not=j\not=k\not=i$, from (\ref{eq:A.1.3}) 
we have that 
\begin{eqnarray}
R_{ijij}\!\!\!\!\!&=\!\!\!\!\!&K(E_i,E_j), \label{eq:A.1.4a} \\
R_{ijkj}\!\!\!\!\!&=\!\!\!\!\!&\frac{1}{2}\Bigl(2K(E_i+E_k,E_j)-K(E_i,E_j)
   -K(E_k,E_j)\Bigr). \label{eq:A.1.4b}
\end{eqnarray}
The number of these components is six, which are in a one-to-one 
correspondence with (appropriate linear combinations of) the six sectional 
curvatures $K(E_i,E_j)$, $K(E_i+E_k,E_j)$. There are no components $R_{ijkl}$ 
in which all the indices, taking the values $1$, $2$ or $3$, would be 
different. 

There are six algebraically independent components of the form $R_{4iji}$, 
$i\not=j$. Explicitly, these are 
\begin{eqnarray}
2R_{4121}\!\!\!\!\!&=\!\!\!\!\!&\bigl(2+2\gamma-\beta^2\bigr)K(E_4
  +E_2,E_1)-K(E_2,E_1)-(1-\beta^2)K(E_4,E_1), \label{eq:A.1.4c} \\
2R_{4131}\!\!\!\!\!&=\!\!\!\!\!&\bigl(2+2\delta-\beta^2\bigr)K(E_4
  +E_3,E_1)-K(E_3,E_1)-(1-\beta^2)K(E_4,E_1), \label{eq:A.1.4d} \\
2R_{4212}\!\!\!\!\!&=\!\!\!\!\!&\bigl(2+2\beta-\gamma^2\bigr)K(E_4
  +E_1,E_2)-K(E_1,E_2)-(1-\gamma^2)K(E_4,E_2), \label{eq:A.1.4e} \\
2R_{4232}\!\!\!\!\!&=\!\!\!\!\!&\bigl(2+2\delta-\gamma^2\bigr)K(E_4
  +E_3,E_2)-K(E_3,E_2)-(1-\gamma^2)K(E_4,E_2), \label{eq:A.1.4f} \\
2R_{4313}\!\!\!\!\!&=\!\!\!\!\!&\bigl(2+2\beta-\delta^2\bigr)K(E_4
  +E_1,E_3)-K(E_1,E_3)-(1-\delta^2)K(E_4,E_3), \label{eq:A.1.4g} \\
2R_{4323}\!\!\!\!\!&=\!\!\!\!\!&\bigl(2+2\gamma-\delta^2\bigr)K(E_4
  +E_2,E_3)-K(E_2,E_3)-(1-\delta^2)K(E_4,E_3). \label{eq:A.1.4h}
\end{eqnarray}
Although $E^\alpha_4$ and $E^\alpha_i$ are spacelike, the 2-planes that they 
span, $[E_4,E_i]$, are not necessarily spacelike. E.g. $[E_4,E_1]$ is 
spacelike precisely when there exists a spacelike vector $X^\alpha$ in this 
2-plane which is orthogonal to $E^\alpha_1$. Thus, let $X^\alpha=aE^\alpha_1+
bE^\alpha_4$, and suppose that $0=g(X,E_1)=-a-b\beta$. Then $a=-b\beta$, and 
hence $g(X,X)=-b^2(1-\beta^2)$. Therefore, $[E_4,E_1]$ is spacelike precisely 
when $\beta^2<1$. In a similar way, $[E_4,E_2]$ is spacelike precisely 
when $\gamma^2<1$; and $[E_4,E_3]$ is spacelike precisely when $\delta^2<1$. 
As one can check easily, these conditions ensure that the vectors $E^\alpha
_4\pm E^\alpha_i$ are all spacelike. A completely similar analysis shows that 
the 2-planes $[E_4+E_1,E_2]$, $[E_4+E_2,E_1]$, $[E_4+E_1,E_3]$, $[E_4+E_3,E_1]$, 
$[E_4+E_2,E_3]$ and $[E_4+E_3,E_2]$ are spacelike if $\gamma^2<2(1+\beta)$, 
$\beta^2<2(1+\gamma)$, $\delta^2<2(1+\beta)$, $\beta^2<2(1+\delta)$, 
$\delta^2<2(1+\gamma)$ and $\gamma^2<2(1+\delta)$, respectively. 

From (\ref{eq:A.1.3}) we have six components of the form $R_{4i4j}$. 
Explicitly: 
\begin{eqnarray}
R_{4141}\!\!\!\!\!&=\!\!\!\!\!&(1-\beta^2)K(E_1,E_4), \label{eq:A.1.4i} \\
R_{4242}\!\!\!\!\!&=\!\!\!\!\!&(1-\gamma^2)K(E_2,E_4), \label{eq:A.1.4j} \\
R_{4343}\!\!\!\!\!&=\!\!\!\!\!&(1-\delta^2)K(E_3,E_4), \label{eq:A.1.4k} \\
2R_{4142}\!\!\!\!\!&=\!\!\!\!\!&\bigl(2-(\beta+\gamma)^2\bigr)K(E_1+E_2,E_4)
  -(1-\beta^2)K(E_1,E_4)-(1-\gamma^2)K(E_2,E_4), \label{eq:A.1.4l} \\
2R_{4143}\!\!\!\!\!&=\!\!\!\!\!&\bigl(2-(\beta+\delta)^2\bigr)K(E_1+E_3,E_4)
  -(1-\beta^2)K(E_1,E_4)-(1-\delta^2)K(E_3,E_4), \label{eq:A.1.4m} \\
2R_{4243}\!\!\!\!\!&=\!\!\!\!\!&\bigl(2-(\gamma+\delta)^2\bigr)K(E_2+E_3,E_4)
  -(1-\gamma^2)K(E_2,E_4)-(1-\delta^2)K(E_3,E_4). \label{eq:A.1.4n}
\end{eqnarray}
Here, three more 2-planes emerged: $[E_1+E_2,E_4]$, $[E_1+E_3,E_4]$ and 
$[E_2+E_3,E_4]$. These are spacelike precisely when $(\beta+\gamma)^2<2$, 
$(\gamma+\delta)^2<2$ and $(\beta+\delta)^2<2$, respectively. 

To express the remaining components of the curvature tensor we should use 
the general formula (\ref{eq:A.1.2}). They are of the form $R_{4ijk}$, where 
$i$, $j$ and $k$ are all different. However, only two of these three, say 
$R_{4123}$ and $R_{4231}$, are independent, because $R_{4123}+R_{4231}+R_{4312}
=0$ holds by the 1st Bianchi identity (which was in fact discovered by 
Ricci \cite{Spivak2}). Explicitly, by (\ref{eq:A.1.2}) they are given by 
\begin{eqnarray}
6R_{4123}\!\!\!\!\!&=\!\!\!\!\!&\bigl(4(1+\gamma)-(\beta+\delta)^2\bigr)K(
  E_2+E_4,E_1+E_3)- \nonumber \\
\!\!\!\!\!&-\!\!\!\!\!&\bigl(4(1+\delta)-(\beta+\gamma)^2\bigr)K(E_1+E_2,
  E_3+E_4)+ \nonumber \\
\!\!\!\!\!&+\!\!\!\!\!&2K(E_1+E_2,E_3)-2K(E_1+E_3,E_2)+K(E_1,E_2)-K(E_1,E_3)+
  \nonumber \\  
\!\!\!\!\!&+\!\!\!\!\!&\bigl(2-(\beta+\gamma)^2\bigr)K(E_1+E_2,E_4)-\bigl(2-
  (\beta+\delta)^2\bigr)K(E_1+E_3,E_4)- \nonumber \\
\!\!\!\!\!&-\!\!\!\!\!&\bigl(2(1+\gamma)-\delta^2\bigr)K(E_2+E_4,E_3)-\bigl(
  2(1+\gamma)-\beta^2\bigr)K(E_2+E_4,E_1)+ \nonumber \\
\!\!\!\!\!&+\!\!\!\!\!&\bigl(2(1+\delta)-\gamma^2\bigr)K(E_3+E_4,E_2)+\bigl(
  2(1+\delta)-\beta^2\bigr)K(E_3+E_4,E_1)+ \nonumber \\
\!\!\!\!\!&+\!\!\!\!\!&(1-\delta^2)K(E_3,E_4)-(1-\gamma^2)K(E_2,E_4), 
   \label{eq:A.1.4o}
\end{eqnarray}
and 
\begin{eqnarray}
6R_{4231}\!\!\!\!\!&=\!\!\!\!\!&\bigl(4(1+\delta)-(\beta+\gamma)^2\bigr)K(
  E_2+E_4,E_1+E_3)- \nonumber \\
\!\!\!\!\!&-\!\!\!\!\!&\bigl(4(1+\beta)-(\gamma+\delta)^2\bigr)K(E_2+E_3,
  E_1+E_4)+ \nonumber \\
\!\!\!\!\!&+\!\!\!\!\!&2K(E_2+E_3,E_1)-2K(E_1+E_2,E_3)+K(E_2,E_3)-K(E_1,E_2)+
  \nonumber \\  
\!\!\!\!\!&+\!\!\!\!\!&\bigl(2-(\gamma+\delta)^2\bigr)K(E_2+E_3,E_4)-\bigl(
  2-(\beta+\gamma)^2\bigr)K(E_1+E_2,E_4)+ \nonumber \\
\!\!\!\!\!&+\!\!\!\!\!&\bigl(2(1+\beta)-\gamma^2\bigr)K(E_1+E_4,E_2)+\bigl(
  2(1+\beta)-\delta^2\bigr)K(E_1+E_4,E_3)- \nonumber \\
\!\!\!\!\!&-\!\!\!\!\!&\bigl(2(1+\delta)-\beta^2\bigr)K(E_3+E_4,E_1)-\bigl(
  2(1+\delta)-\gamma^2\bigr)K(E_3+E_4,E_2)- \nonumber \\
\!\!\!\!\!&-\!\!\!\!\!&(1-\beta^2)K(E_1,E_4)-(1-\delta^2)K(E_3,E_4).
   \label{eq:A.1.4p}
\end{eqnarray}
In these expressions, three more 2-planes appeared: $[E_4+E_1,E_2+E_3]$, 
$[E_4+E_2,E_1+E_3]$ and  $[E_4+E_3,E_1+E_2]$. These are spacelike precisely 
when $(\gamma+\delta)^2<4(1+\beta)$, $(\beta+\delta)^2<4(1+\gamma)$ and 
$(\beta+\gamma)^2<4(1+\delta)$, respectively. 

Thus, to summarize, the conditions under which the 2-planes in the 
expressions of the independent components of the curvature tensor in the 
basis $\{E^\alpha_{\ua}\}$ are spacelike are 
\begin{eqnarray*}
&{}& \beta^2,\,\gamma^2,\,\delta^2<1; \hskip 35pt (\beta+\gamma)^2,\,
  (\gamma+\delta)^2,\,(\delta+\beta)^2<2; \\
&{}& \gamma^2,\,\delta^2<2(1+\beta); \hskip 20pt \delta^2,\, \beta^2<2
  (1+\gamma),\, \hskip 20pt \beta^2,\,\gamma^2<2(1+\delta); \\
&{}& (\beta+\gamma)^2<4(1+\delta),\, \hskip 10pt (\gamma+\delta)^2<4(1+\beta),
  \, \hskip 10pt (\delta+\beta)^2<4(1+\gamma);
\end{eqnarray*}
where $\beta^2+\gamma^2+\delta^2=1+\alpha^2$ and $\alpha>0$. Since it does not 
seem natural to distinguish a priori any spatial direction in the definition 
of $E^\alpha_4$, we can try to satisfy these conditions with $\beta=\gamma=
\delta>0$, which would yield $\beta^2=(1+\alpha^2)/3$. As it can be checked 
directly, with this choice all the conditions above can, in fact, be 
satisfied if $1/3<\beta^2<1/2$, i.e. if $0<\alpha<1/\sqrt{2}$. Thus we 
assume that $E^\alpha_4$ has this structure. 

The resulting these 21 spacelike 2-planes can be memorized easily by the 
following picture. Let us form a tetrahedron whose vertices are labeled by 
${\ua}=1,2,3,4$, and the midpoint of the edge between the adjacent vertices 
${\ua}$ and ${\ub}$ by the unordered pair ${\ua}{\ub}$. Then six of the 
spacelike 2-planes are represented by the six edges, twelve of the 2-planes 
by straight line segments between the midpoints and the not adjacent 
vertices, and the remaining three 2-planes by straight line segments between 
the midpoints on the not adjacent edges. 

As we have already noted, since in four dimensions the number of the 
algebraically independent components of the curvature tensor is twenty and 
the curvature tensor defines the sectional curvatures, only 20 of the 21 
sectional curvatures can be considered to be independent. By 
(\ref{eq:A.1.4a})-(\ref{eq:A.1.4n}) the 18 algebraically independent 
components of the curvature tensor can be expressed by the 18 sectional 
curvatures $K(E_i,E_j)$, $K(E_4,E_i)$, $K(E_i+E_j,E_k)$, $K(E_4+E_i,E_j)$ and 
$K(E_4,E_i+E_j)$. Hence these are independent. Thus, only two of the 
remaining three sectional curvatures, $K(E_4+E_1,E_2+E_3)$, $K(E_4+E_2,
E_1+E_3)$ and $K(E_3+E_3,E_1+E_2)$, can be independent. In fact, e.g. a 
direct calculation shows that their sum is a linear combination of the 18 
independent sectional curvatures, viz. that the identity 
\begin{eqnarray}
(1+\beta-\beta^2)\Bigl(\!\!\!\!\!&{}\!\!\!\!\!&K(E_4+E_1,E_2+E_3)+
  K(E_4+E_2,E_1+E_3)+K(E_4+E_3,E_1+E_2)\Bigr)= \nonumber \\
=\!\!\!\!\!&{}\!\!\!\!\!&-\frac{1}{2}\Bigl(K(E_1,E_2)+K(E_1,E_3)+K(E_2,E_3)
  \Bigr)+ \nonumber \\
\!\!\!\!\!&{}\!\!\!\!\!&+\frac{1}{2}\Bigl(K(E_1+E_2,E_3)+K(E_1+E_3,E_2)+
  K(E_2+E_3,E_1)\Bigr)- \nonumber \\
\!\!\!\!\!&{}\!\!\!\!\!&-\frac{1}{2}(1-\beta^2)\Bigl(K(E_4,E_1)+K(E_4,E_2)+
  K(E_4,E_3)\Bigr)+ \nonumber \\
\!\!\!\!\!&{}\!\!\!\!\!&+\frac{1}{2}(1-2\beta^2)\Bigl(K(E_4,E_1+E_2)+
  K(E_4,E_1+E_3)+K(E_4,E_2+E_3)\Bigr)+ \nonumber \\
\!\!\!\!\!&{}\!\!\!\!\!&+\frac{1}{4}(2+2\beta-\beta^2)\Bigl(K(E_4+E_1,E_2)+
  K(E_4+E_1,E_3)+K(E_4+E_2,E_1)+ \nonumber \\
\!\!\!\!\!&{}\!\!\!\!\!& \hskip 70pt +K(E_4+E_2,E_3)+K(E_4+E_3,E_1)+
  K(E_4+E_3,E_2)\Bigr) \label{eq:A.1.5}
\end{eqnarray}
must hold. The sectional curvature corresponding to any other non-degenerate 
2-plane at $q$ must be a linear combination of these twenty independent 
ones. Indeed, if $X^\alpha,Y^\alpha\in T_qM$ are unit spacelike vectors that 
are orthogonal to each other, then $K(X,Y)=R_{\ua\ub\uc\ud}X^{\ua}Y^{\ub}X^{\uc}
Y^{\ud}$, where the components of the curvature tensor are linear combinations 
of the 20 independent sectional curvatures via (\ref{eq:A.1.2}), and hence 
$K(X,Y)$ will also be a linear combination of the 20 independent sectional 
curvatures. 

If the number of the algebraically independent components of the curvature 
tensor is less than twenty, e.g. when the Ricci part of the curvature 
vanishes, then the number of the independent sectional curvatures is also 
less. The vanishing of the whole curvature at $q$ is, of course, equivalent 
to the vanishing of all the sectional curvatures.


\end{document}